\documentclass[english,format=acmsmall, review=false, screen=true]{acmart}
\usepackage[T1]{fontenc}
\usepackage[latin9]{inputenc}
\PassOptionsToPackage{ruled}{algorithm2e}
\usepackage{color}
\usepackage{array}
\usepackage{textcomp}
\usepackage{multirow}
\usepackage{algorithm2e}
\usepackage{amsbsy}
\usepackage{graphicx}

\makeatletter
\newcommand*{\rom}[1]{\expandafter\@slowromancap\romannumeral #1@}
\makeatother

\makeatletter

\acmJournal{TOMM}

\acmYear{2020}
\acmMonth{3}
\copyrightyear{2020}
\setcopyright{acmlicensed}
\received{Oct 2019}
\received[revised]{Jan 2020}
\received[accepted]{March 2020}
\acmDOI{10.1145/3387164}
\providecommand{\tabularnewline}{\\}

\newcommand{\redefineshortauthors}[1]{\renewcommand{\shortauthors}{#1}}

\SetAlFnt{\small}
\SetAlCapFnt{\small}
\SetAlCapNameFnt{\small}
\SetAlCapHSkip{0pt}
\IncMargin{-\parindent}
\makeatother
\usepackage{babel}

\begin{document}

\title{Deep Triplet Neural Networks with Cluster-CCA for Audio-Visual Cross-Modal Retrieval}
\author{Donghuo Zeng}
\orcid{0000-0002-0294-6620}
\affiliation{\institution{National Institute of Informatics, SOKENDAI}\streetaddress{2-1-2 Hitotsubashi}\city{Chiyoda-ku}\state{Tokyo}\postcode{100-0003}\country{Japan}}

\author{Yi Yu}
\affiliation{\institution{National Institute of Informatics, SOKENDAI}\streetaddress{2-1-2 Hitotsubashi}\city{Chiyoda-ku}\state{Tokyo}\postcode{100-0003}\country{Japan}}

\author{Keizo Oyama}
\affiliation{\institution{National Institute of Informatics, SOKENDAI}\streetaddress{2-1-2 Hitotsubashi}\city{Chiyoda-ku}\state{Tokyo}\postcode{100-0003}\country{Japan}}

\renewcommand{\abstractname}{Acknowledgements}
\begin{abstract}
Cross-modal retrieval aims to retrieve data in one modality by a query in another modality, which has been a very interesting research issue in the field of multimedia, information retrieval, and computer vision, and database. 
Most existing works focus on cross-modal retrieval between text-image, text-video, and lyrics-audio. Little research addresses cross-modal retrieval between audio and video due to limited audio-video paired datasets and semantic information. The main challenge of audio-visual cross-modal retrieval task focuses on learning joint embeddings from a shared subspace for computing the similarity across different modalities, where generating new representations is to maximize the correlation between audio and visual modalities space. In this work, we propose a novel deep triplet neural network with cluster canonical correlation analysis (TNN-C-CCA), which is an end-to-end supervised learning architecture with audio branch and video branch. We not only consider the matching pairs in the common space but also compute the mismatching pairs when maximizing the correlation. In particular, two significant contributions are made: i) a better representation by constructing deep triplet neural network with triplet loss for optimal projections can be generated to maximize correlation in the shared subspace. ii) positive examples and negative examples are used in the learning stage to improve the capability of embedding learning between audio and video. Our experiment is run over 5-fold cross-validation, where average performance is applied to demonstrate the performance of audio-video cross-modal retrieval. The experimental results achieved on two different audio-visual datasets show the proposed learning architecture with two branches outperforms existing six CCA-based methods and four state-of-the-art based cross-modal retrieval methods.
\end{abstract}

\selectlanguage{english}%

\maketitle
\redefineshortauthors{DH. Zeng et al.}
\section{Introduction}
\begin{figure}[t]
\includegraphics[width=10cm,height=5cm]{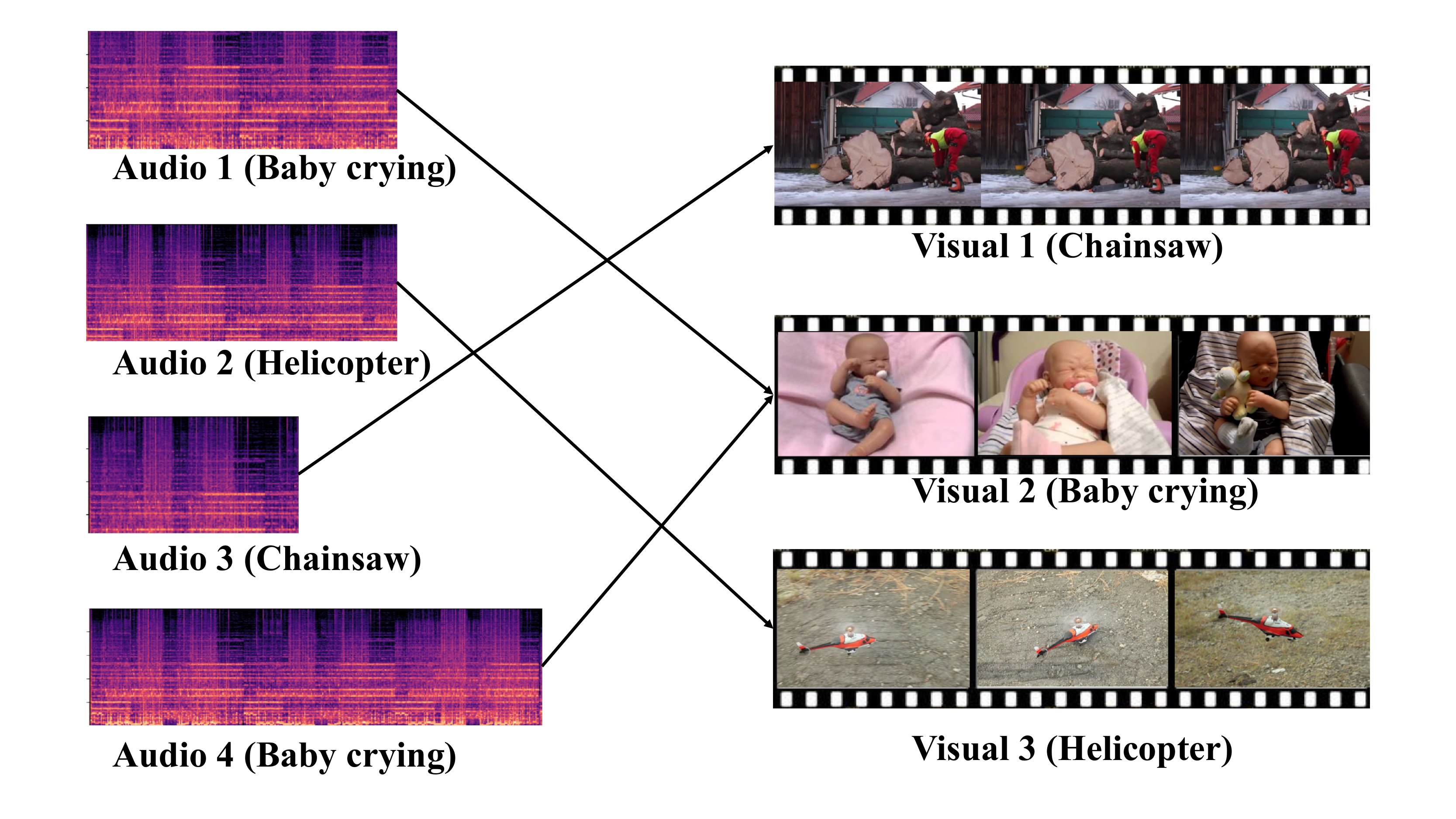}
\centering
\caption{The overview of audio-to-visual cross-modal retrieval. Given an audio to find similar visual contents, the matched visual contents which have the same semantic label as an audio query. The audio query can be any kind of arbitrary length.}
\label{overall_retrieval}
\end{figure}
With the high-speed development of innovative technology and user interaction on the Internet, various multimedia data and information have been aggregated. It results in a heterogeneous gap between different modality data, which brings a big challenge for efficiently and effectively cross-modal retrieval between data from different modalities. 
In the past, researches have focused on building the similarity link between every two data points from different modalities for cross-modal retrieval tasks, which has made big successes in cross-modal retrieval, such as image-text~\cite{wang2015image, wang2016learning}, audio-text~\cite{yu2019deep}, and video-text~\cite{sivic2003video}. In particular, the visual and auditory senses of human being are the most important ways to understand the living environment and understand the world, seen in Fig.~\ref{overall_retrieval}. For instance, when hearing a helicopter sound, a helicopter can be imagined in your mind. When you see lightning, subconsciously the thunder is coming soon. Unfortunately, due to the limited audio-video paired dataset and semantic category information, little research works on audio-visual cross-modal retrieval~\cite{shah2014advisor}. This motivates us to mimic the mutual-aid based learning process and extract cognitive patterns from human being.

Cross-modal retrieval between every two data points from different modalities has a challenge of the heterogeneous gap of data structure
among the modalities, which requires us to formulate a joint representation space, where the similarity of different data modalities reflects the semantic closeness between their corresponding original inputs by correlation learning.

Recently, most methods for correlation learning are to bridge the gap of different modalities by learning joint embedding or representation learning, which has achieved great success in cross-modal retrieval tasks~\cite{RanjanRJ15,hardoon2004canonical, peng2016cross, wang2013learning}. The purpose of representation learning is to find projections of data examples from a different data set into a shared subspace, where the similarity relationship between them can be measured. 

The typical representation learning method CCA~\cite{hardoon2004canonical} is to find linear transformations of two-view of data as inputs via maximizing the pairwise correlation. However, if there is a nonlinear relation between two instances, CCA has no capability to always extract useful features. Kernel-CCA~\cite{akaho2006kernel} uses the kernel method to CCA, which enables the nonlinear transformation for two-view of data. With the rapid growth of deep neural network (DNN) techniques, the DNN model has been progressively applied in cross-modal retrieval tasks~\cite{andrew2013deep, feng2014cross, ngiam2011multimodal, yan2015deep}. For example, Deep Canonical Correlation Analysis (DCCA)~\cite{andrew2013deep}, which is used for learning complex nonlinear transformations of the different datasets. DCCA can learn nonlinear transformations without the inner product computation of Kernel-CCA. Also, DCCA has no hyper-parameters limited in the representation learning unlike kernel-CCA limited in the fixed kernel. The current cross-modal retrieval model also tries to keep the pairwise correlation with the joint predefined semantic categories, where each category contains many pairs of cross-modal data. CCA, Kernel-CCA and DCCA cross-modal retrieval methods focus on the pairwise correlation only. However, the different samples with the same category convey the same semantic information which might be neglected. In theory, to solve this issue, it requires a model that can preserve all the semantic information during the representation learning, where the heterogeneous gap in the pairwise samples is minimized while non-pairwise samples with the same semantic categories are maximized.

Cluster-CCA~\cite{rasiwasia2014cluster}
can preserve all the semantic information by applying a one-to-one correspondence between all pairs from the cross-modal dataset and use standard CCA to learn the projections. Cluster-CCA can learn joint representations that maximize the correlations between the two different modalities and segregating the different categories in the shared subspace. Cluster-CCA tries to enhance the similarity inside the category between data from different modalities. Inspired by Cluster-CCA and DCCA, an improved C-DCCA\cite{yu2018category} is proposed to learn the nonlinear correlation between data from different modalities and simultaneously consider the similarity within the category across modality data.  However, the above methods cannot guarantee all the similarity distance of two instances from different modalities of the same category is similar than that of two instances from different modalities of the different categories. 

To settle this problem, it needs to completely consider all the positions of data points in the common space. The previous joint embedding learning methods, after the two branch networks are optimized, the CCA projections are calculated only one time. It is impossible to completely focus on the distribution of all the data points in the shared subspace. 

To figure out this issue, our first contribution is that deep TNN is proposed to maximize the correlation between every two instances from different modalities with the same category while minimizing the correlation between data from different modalities from different categories during training. In other words, each data point from one modality is more close to samples with the same semantic category from the other modality (namely positive samples). Simultaneously, the data point is farther from instances with different categories. (namely negative samples). The deep TNN used here is to apply deep neural networks with backpropagating errors and use triplet loss to update the weights of the neural network during the training. The second contribution is that all the data points within a batch size is considered to meet storage limitation instead of using all the position of data points space. Finally, our architecture is evaluated on two video datasets. MV-10K dataset is selected from the YouTube-8M video dataset by us, which is utilized in our previous work~\cite{zeng2018audio}.
To evaluate the extendability of our algorithms, VEGAS dataset~\cite{zhou2018visual} is used in the experiments. The experimental results demonstrate that the proposed embedding learning  architecture significantly surpasses the existing six CCA-based methods and four state-of-the-art methods in cross-modal retrieval.

The rest of this paper is organized as follows. In Section 2, we show  some works related to our proposed approach. Section 3 presents our proposed TNN-C-CCA model. Section 4 introduces the experimental results of our approach achieved on two audio-visual cross-modal datasets with the analyses. Finally, Section 5 makes a conclusion of this paper.

\section{Related Work}
In this section, we briefly present two main research lines related to our method, including cross-modal retrieval and triplet neural networks.
\subsection{Cross-modal Retrieval} Different from retrieval in the same modality, such as image retrieval~\cite{WangLHT19}, cross-modal retrieval is used for implementing a retrieval task across different modalities. such as image-text\cite{karpathy2014deep, wang2015image, yan2015deep, wang2016learning}, video-text\cite{sivic2003video}, and audio-text\cite{yu2019deep} cross-modal retrieval. The main challenge of cross-modal retrieval is the modality gap and the key solution of cross-modal retrieval is learning joint embedding for different modalities. Learning joint embedding is not only a solution of cross-modal retrieval and also applied for other multimedia tasks, such as image classification~\cite{YanNLGYX16}, video question and answering~\cite{Zhu2017}. As for our task, cross-modal retrieval aim at generating new representations from different modalities in the shared subspace, such that new generated features can be applied in the computation of distance metrics, such as Cosine distance and Euclidean distance. Generally, the output space of the architecture after training can be real-valued common space and binary representation learning space. 

\subsubsection{Real-valued common space}\mbox{}

In this paper, we focus on common space learned by the real-valued representation learning for cross-modal retrieval. This kind of representation learning can be classified into three categories as follows.

\paragraph{Unsupervised methods}
Canonical correlation analysis (CCA) is one of the most popular cross-modal embedding models, which aims at finding a pair of linear transformations to maximize the correlation between two different modalities. 
The work~\cite{rasiwasia2010new} uses CCA to calculate the cross-modal correlations between image and text. A novel method for a cross-modal association is Cross-modal Factor Analysis (CFA)~\cite{li2003multimedia}, which is used for audio-image cross-modal retrieval task. Another way to reduce the dimension is to find unified feature subspace by a principle of collective component analysis (CoCA)~\cite{shi2012dimensionality}, where two different modality data points should be correspondence among the projections in the shared subspace and the similarity between the paired data points should be maximized. The paper~\cite{wang2015image} proposed a cross-modal projection matching (CMPM) loss and projection classification loss for learning the discernible different embeddings space. The CMPM loss minimizes the KL divergence between each pair and groups the new representations into different clusters. 

\paragraph{Semi-supervised methods}
Different from the unsupervised methods computing the correlation of projections without any category information during the training, semi-supervised methods solve the problem of limited labels dataset. The approach MVML-GL\cite{xu2013survey} with semi-supervised learning is to reveal the latent feature space by keeping global consistency structure and local geometric architecture.  The paper\cite{socher2010connecting} proposed a semi-supervised model which is suitable for few labeled images and large unaligned textual documents to locate image regions to texts. GSS-SL\cite{zhang2017generalized} is a semi-unsupervised method, which is via predicting more related labels for unlabeled data with label graph constraint and the labels directly are regarded as the semantic information of multimedia data. The paper~\cite{YangNXLZP12} proposed LRGA ranking method and long-term RF algorithm to learn multimedia data representation by exploiting the history of RF information and multimedia data distribution by users.

\paragraph{Supervised methods}
During the correlation learning of cross-modal retrieval, the semantic class information can be applied for the similarity learning. GAM~\cite{sharma2012generalized} is an extension of the canonical correlation analysis method, which focuses on keeping the function of popular supervised and unsupervised feature extraction approaches by developing a quadratic program to get a single nonlinear subspace over different feature spaces.
This paper~\cite{karpathy2014deep} proposed a model focusing on a finer level and fragment of both images and sentences in the shared subspace.
In this paper~\cite{wang2015deep}, they utilized deep CNN features as the visual inputs and topic features as textual semantic inputs. They proposed a regularized deep neural network for nonlinear semantic correlation across modalities. They put forward the intra-modal regularization to learn joint embeddings with Intra- and Inter- modal relation.
Topic modeling on Latent Dirichlet Allocation (LDA) is always applied in multimodal data, especially for the Document Neural Autoregressive Distribution Estimator (DocNADE) which is the best for document modeling. In this paper~\cite{zheng2014topic}, they put forward a method called SupDocNADE which is a supervised improvement method of DocNADE. This method enhances the discriminative power of topic features. CM-GANs~\cite{peng2019cm} learns discriminative joint embedding to bridge the heterogeneity gap. Zero-shot can be regarded as an extension of supervised learning, the TANSS~\cite{8771379} model applied zero-shot to obtain a self-supervised semantic sub-network that can enhance sub-network generalization for unseen labels.

\subsubsection{Binary representation learning space}\mbox{}

As the huge multimodal data increase on the Internet, it is becoming difficult to satisfy the requirement of storage and retrieval capabilities over a big cross-modal multimedia dataset. A solution to speeding up the search is cross-modal hashing~\cite{DBLPRastegariCFHD13, IrieAT15}, which is widely employed to learn binary representations in the common space. Cross-modal hashing methods try to learn correlations between two modalities of data, which can improve the accuracy of cross-modal search by projecting the data into a common Hamming space. However, hashing code generated by traditional methods~\cite{DBLPRastegariCFHD13} is less discriminative to different categories. The DCH~\cite{xu2017learning} model directly learns discriminative hashing code by discriminative classification.
The MDBE~\cite{wang2016multimodal} model learns discriminative hash code by preserving discriminability with classification method and keeping similarity with shared structure among the data from different modalities. The CRE~\cite{Hu0S0HS19} model proposes different modality-specific models to bridge heterogeneity gap, meanwhile, projecting different modalities of data into Hamming space to reconstruct embedding, which reduces the optimization complexity and preserves the inter-category information. The paper ~\cite{XuHSTL16} aims at overcoming the loss information of quantization in hashing code learning, by rotating the joint space to produce better unified hashing codes.

Using common space learning methods usually to preserve the characteristics of original feature explicitly is a challenging issue. Attention mechanism in DNN ~\cite{peng2018modality} is used to construct modality-specific semantic space, which is directly generated without common space learning. Adversarial learning based model DAML~\cite{xu2019deep} not only learns latent feature subspace through minimizing the intra-category correlations and maximizing the inter-category correlations, but also proposed a modality classifier to guarantee the output embedding is also statistically indistinguishable. 

\subsection{Triplet Neural Network}
Triplet neural network model is the extension of the Siamese Neural Network (SNN)~\cite{koch2015siamese}, SNN has typically succeeded in object tracking~\cite{bertinetto2016fully}, image recognition~\cite{koch2015siamese}, person re-identification~\cite{hermans2017defense,ShenJCJCQH19}, and face recognition ~\cite{schroff2015facenet,JiHYSS19} tasks. SNN consists of twin networks, and the network is symmetric, which tries to employ a unique network to rank similarity between two different modality inputs. This system can generate discriminative features and enhance the generalization power of the network. Triplet neural network consists of three networks, where the weights of them are updated by the triplet loss. During the training, the triplet loss requires that these anchor samples are more similar to positive samples than to negative samples by a hyper-parameter margin.

Some ranking based approaches, it is effective for the representation learning in the cross-modal retrieval tasks. ACMR~\cite{wang2017adversarial} model applies triplet loss method to keep the intra-modal distinguishable and inter-modal unchangeable in the feature projector, which minimizes the gap between all samples of different modalities with the same semantic labels, in the meanwhile, maximizing the distances of samples belonging to different labels. Methods with triplet loss in cross-modal retrieval can better preserve the representation structure of images and texts when projected into the shared subspace by means of the triplet constraint in the new representation generation by triplet loss. 
Deep image-text embedding~\cite{wang2016learning} is trained by applying a large margin objective function with cross-modal ranking constraints. Instead of computing on all triplets, they sampled triplets in each mini-batch by selecting top K most hard matches according to the similarity distance and exploit Stochastic Gradient Descent (SGD) to minimize the loss function. The average of the loss of all batches is used.  
Learning ranking functions during the training of cross-modal retrieval gets popular in recent researches, which has been  regarded as a fundamental problem. In this paper~\cite{yao2015learning}, they proposed a ranking canonical correlation analysis (RCCA) method to solve the limitation of two paradigms. The image vector model highly requires the quality of textual description and the manual labeling is hard to obtain. Through establishing triplet from click-through data to optimize the objective function, the triplet consists of a query, a higher clicked image, and a lower clicked image. In order to preserve the relations in the triplets, they build a triplet loss to optimize the weights.
To better understand and model the networks for connecting the information of cross-views in the shared spaces, the proposed structure-preserving metric model (SPML)~\cite{shaw2011learning} learns a Mahalanobis distance metric so that the structure of the network is preserved
by a hinge-loss over triplets, which consists of a node, its non-neighbor node, and its neighbor node. They apply stochastic subgradient descent to optimize the triplet loss function and sample a batch of triplets to train. 

Nevertheless, these models for learning embeddings to preserve the joint relation of examples are not designed for audio-visual cross-modal retrieval. In this work, we propose a deep triplet neural network to optimize the embedding generated from linear cluster-CCA and simultaneously solve the problem that the distance of similar samples is farther than the distance of dissimilar samples during the training.
\begin{table}[h!]
  \begin{center}
    \caption{Configuration of TNN-C-CCA}
    \label{config:tnn-c-cca}
    \begin{tabular}{r|l} 
    \hline
       log mel-spectrogram audio inputs& 96x64 \\
       Output of visual branch& L\footnotemark *1024 \\
       Output of audio branch& L*128 \\
       Output of Cluster-CCA& 10 \\
       Fully connected layers for audio& [100, 100, 100, 10] \\
       Fully connected layers for visual& [200, 200, 200, 10] \\
       Output of TNN-C-CCA& 10 \\
      \hline
    \end{tabular}
  \end{center}
\end{table}
\footnotetext{L is the number of frames in a video, by decoding each video at one frame per second.}

\begin{figure}[t]
\includegraphics[width=14cm, height=7.5cm]{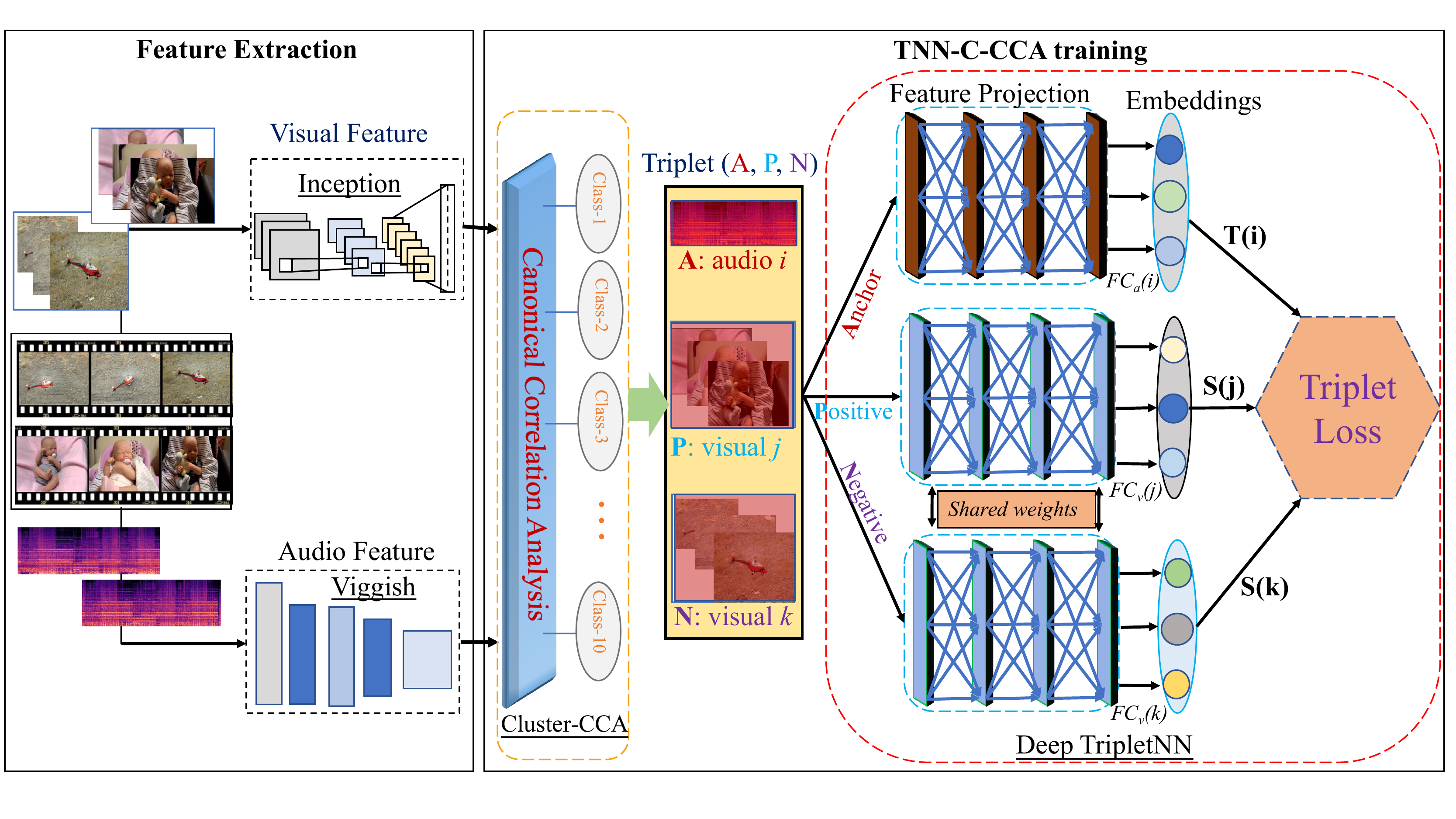}
\centering
\caption{The overall framework of our TNN-C-CCA model. It consists of two parts: feature extraction and TNN-C-CCA training. We apply Inception V3 and Vggish model to extract feature, then explore cluster-CCA to learn the correlation with cluster segregating and select triplets as input for deep TNN training. In the deep TNN, there are three branches: anchor, positive, and negative. Positive and negative branches shared the same weights. Anchor branch is trained by audio data, positive and negative branches are trained by visual data. The detailed description is shown in section 3.3.}
\label{architecture}
\end{figure}
\section{Architecture}
Our deep architecture generally can be divided into two different parts: feature extraction and TNN-C-CCA training, as shown in Fig.~\ref{architecture}. The configuration of TNN-C-CCA used in this work is shown in Table~\ref{config:tnn-c-cca}. Outputs of visual branch and audio branch respectively are 1024-dimensional and 128-dimensional, which are mapped to 10-dimensional by cluster-CCA. Deep triplet neural network consists of 4 fully connected layers respectively for audio embedding and visual embedding and outputs a feature vector with a size of 10. The motivation of our architecture is to take advantage of the two models. Cluster-CCA is to establish one-to-one correspondences between all possible pairs by given categories information across the two modalities to maximize the correlation between the latent representation of two different modalities via CCA. 
deep TNN aims to enforce the relation of similar samples and simultaneously weaken the relation of dissimilar samples.
Particularly, using more negative samples and positive samples during the training of Deep TNN improves the discriminative capability of the embedding space.

\subsection{Distance Metric}
In previous face identity tasks~\cite{schroff2015facenet, hermans2017defense}, they used Euclidean distance $D(T(i)-T(j)) = ||T(i)-T(j)||_{2}^{2}$ to calculate the distance between the image anchor and the positive image or the negative image, where $i$ and $j$ are from the same modality of image. However, in our experiment, we apply a cosine similarity for the final representation comparison at the end of the whole architecture. Our distance metric can be defined as follows:
\begin{equation}
\begin{aligned}
||x, y||_{cosine-distance} = 1 - \frac{\sum_{k=1}^{n} x_{k}y_{k}}{\sqrt{\sum_{k=1}^{n} x_{k}}\sqrt{\sum_{k=1}^{n} y_{k}}},
\end{aligned}
\end{equation}

where $n$ is the dimension of vector $x$ and $y$, its iteration $k$ ranges from 1 to $n$. The scale of the Cosine distance ranges from 0 to 2 and the effective margin shares the same scale, normally it is set to 0.5.

\subsection{Input Feature Representation}
The audio feature is extracted by a pre-trained VGGish model with the Mel spectrogram feature as input. We apply the librosa\footnote{https://librosa.github.io/librosa/} library to achieve Mel spectrogram feature extraction with default parameters: hop size=512, nftt=2,048. We use mel-frequency cepstrum (MFC) to do sound processing by applying linear cosine transform of a log power spectrum to represent the short term power spectrum of audio. 

Deep learning pre-trained audio feature extractors can be divided into two categories: training with audio label and training without audio label.

We choose the VGGish model training with audio label which can capture the label information in the final feature obtained and is suitable for our dataset with labels. VGGish model is a VGG-like model, which is trained on a large-scale dataset named AudioSet for the audio classification task. Compared with the VGG model, the VGGish model changes the input size and cuts the last group of convolutional layers and max pool layers, and uses a 128-wide fully connection layer at the end of the neural network. The inputs of audio features are re-sampled to 16kHZ. The window size of FFT is 25ms, with a window hop of 10ms.

VGGish model converts the audio feature inputs into 128-D semantic high-level feature, which is used for audio-visual cross-modal retrieval.

Visual feature is extracted by the state-of-the-art deep CNN model named Inception V3~\cite{szegedy2016rethinking, szegedy2017inception}. The Inception V3 model ~\footnote{https://github.com/google/youtube-8m/tree/maste /feature\_extractor} is popularly-used in the image recognition task which can reach high accuracy on the ImageNet dataset~\cite{ioffe2015batch, deng2009imagenet}. Recently, the pre-trained Inception V3 model is exploited as a video feature extractor~\cite{abu2016youtube}. The output of the Inception V3 model is frame-level features along with the input of pre-processing videos.  By decoding each video at one frame per second, these decoded videos are fed to the Inception V3 network and adopt the ReLU activation in the last hidden layer before the prediction layer. The feature representation is 2,048 dimensional per frame of videos and keeps the front 360 frames. After that, PCA approach is applied to reduce the dimensions per frame to 1,024 by using the quantization method. 

Finally, we apply the general semantic audio and visual features as the input of our TNN-C-CCA architecture, the input vector is the global average of frame-level feature which is computed by the number of frames, so that the input audio feature is 128 dimensions and the visual feature is 1024 dimensions.

\subsection{Cluster-CCA}
CCA is used for exploring the relationship between two multivariate sets of vectors, such as $x\in R^{A}$ and $y\in R^{B}$ with zero-mean, and the pair format is like ($x_{i}$, $y_{i}$). The goal of CCA is to find a new coordinate for $x$ and $y$ by direction $w\in R^{A}$ and $u\in R^{B}$ respectively, such that the correlation between these two sets is maximized. The correlation can be defined as follows:
\begin{equation}
\begin{aligned}
corr = \frac{w^{'}C_{xy}u}{\sqrt{w^{'}C_{xx}w}\sqrt{u^{'}C_{yy}u}},
\label{cca1}
\end{aligned}
\end{equation}

\begin{equation}
\begin{aligned}
C_{xx} = E[xx^{T}] = \frac{1}{n}\sum_{i=1}^{n} x_{i}x_{i}^{T}, \quad
C_{yy} = E[yy^{T}] = \frac{1}{n}\sum_{i=1}^{n} y_{i}y_{i}^{T}, \quad
C_{xy} = E[xy^{T}] = \frac{1}{n}\sum_{i=1}^{n} x_{i}y_{i}^{T}, 
\label{cca2}
\end{aligned}
\end{equation}

Where $corr$ is the correlation, $C_{xx}$, and $C_{yy}$ are the co-variance metrics, $C_{xy}$ is the cross-variance metrics. Here $E(*)$ is the expectation function. Normally, the problem is regarded as an eigenvalue problem, suppose $w$ is the top eigenvector, the problem can be represented as follows:
\begin{equation}
\begin{aligned}
C_{xx}^{-1}C_{xy}C_{yy}^{-1}C_{yx}w = \lambda^{2} w,
\label{eigenvalues}
\end{aligned}
\end{equation}

CCA has been successfully applied to several multimedia problems, such as cross-modal retrieval. However, CCA is suitable for calculating pairwise correlation similarity from different modalities and not available for calculating correlation similarity within a cluster. CCA will be ineffective for learning representation with a cluster in this case.
Cluster-CCA is a variant of CCA~\cite{hardoon2004canonical} with consideration of the cluster segregating by establishing one-to-one correspondences from all pairs of data points in a given cluster across the two different modalities, then apply CCA to learn the projections. 
\begin{equation}
\begin{aligned}
corr = \frac{w^{'}C_{xy}^{'}u}{\sqrt{w^{'}C_{xx}^{'}w}\sqrt{u^{'}C_{yy}^{'}u}},
\label{ccca_1}
\end{aligned}
\end{equation}

The three types of variances can be formulated as follows:
\begin{equation}
\begin{aligned}
C_{xx}^{'} = \frac{1}{L}\sum_{c=1}^{C}\sum_{i=1}^{|X_{c}|} |Y_{c}|x_{i}^{c}x_{i}^{cT},  \quad 
C_{yy}^{'} = \frac{1}{L}\sum_{c=1}^{C}\sum_{j=1}^{|Y_{c}|} |X_{c}|y_{j}^{c}y_{j}^{cT},  \quad
C_{xy}^{'} = \frac{1}{L}\sum_{c=1}^{C}\sum_{i=1}^{|X_{c}|}\sum_{j=1}^{|Y_{c}|} x_{i}y_{j}^{cT},
\label{ccca_2}
\end{aligned}
\end{equation}

Where $L = \sum_{c=1}^{C} |X_{c}||Y_{c}|$ is the sum number of all pairs. Similar to CCA, the optimization problem can be regarded as an eigenvalue problem like formulation~(\ref{eigenvalues}). Here we assume that the covariance is calculated for the zero-mean
random variables.

\subsection{Deep Triplet Neural Network}
\begin{figure}[t]
\includegraphics[width=14cm,height=4cm]{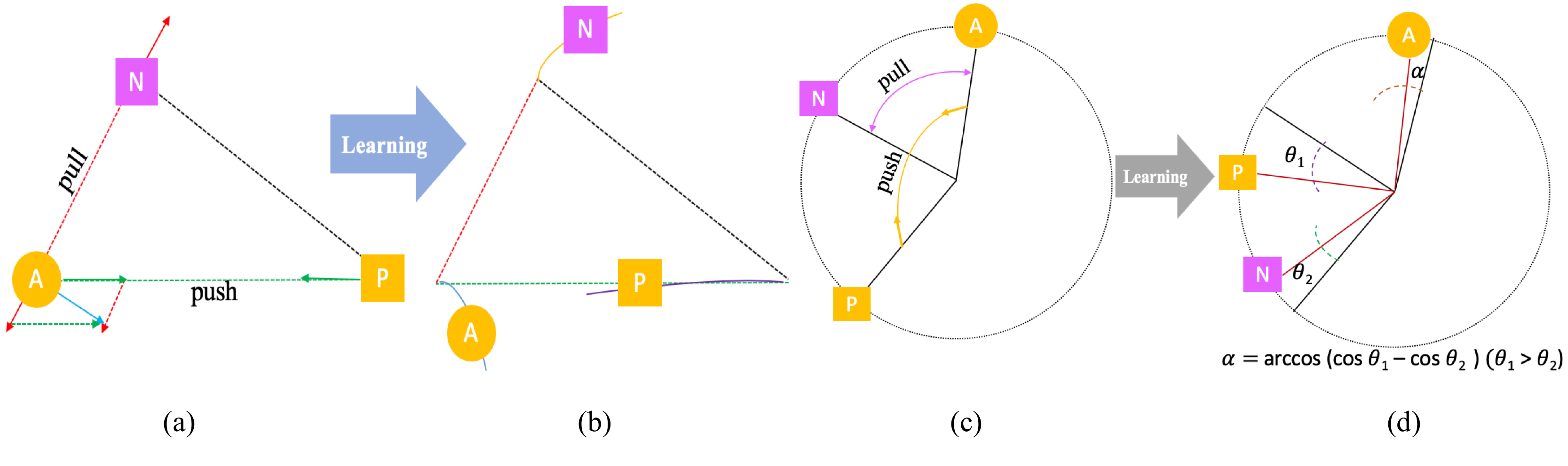}
\centering
\caption{
(a) and (b) show the traditional triplet loss minimizes the Euclidean distance between (anchor, positive) and (anchor, negative) with a fixed margin and optimal gradient back-propagation direction; (c) and (d) present our triplet loss through minimizing the Cosine distance between (anchor, positive) and (anchor, negative) with fixed margin and optimal gradient back-propagation direction.}
\label{TNNloss}
\end{figure}

The Deep Triplet Neural Network is an end-to-end training, as shown in Fig.~\ref{architecture}, which is optimized by triplet loss~\cite{schroff2015facenet} at the end of cross-modal retrieval architecture. For example, in audio-to-visual retrieval process, we try to obtain an audio $i$ represented by $T(i)$ and a visual $j$ represented by $S(j)$, a visual $k(k\neq i)$ represented by $S(k)$, where T(.) and S(.) are the output of Cluster-CCA model, i and j from the same category, i and k from different categories.
Here we want to guarantee audio sample i (Anchor) of one specific category is closer to visual sample j (Positive) of the same category than any visual sample k (Negative) of any other category.
As shown in Fig.~\ref{TNNloss}.
Triplet loss will pull Anchor and Positive samples, simultaneously push Anchor and Negative samples. The condition is represented as follows.
\begin{equation}
\begin{aligned}
||T(i)-S(j)||_{cosine-distance}+\alpha < ||T(i)-S(k)||_{cosine-distance}, \\
Lab^{i} = Lab^{j},\quad Lab^{i} \neq Lab^{k} (i \neq k), \quad
\forall (i, j, k) \in \Lambda,
\label{triplet}
\end{aligned}
\end{equation}

Where $\alpha$ is a margin that is used for reinforcing the Cosine distance among anchor, positive and negative. $\Lambda$ is the collection of all possible triplets in the training dataset.
The triplet loss can be defined as follows:

\begin{equation}
\begin{aligned}
Loss = Max\{\sum_{i}^{N} [||T(i)-S(j)||_{cosine-distance} - ||T(i)-S(k)||_{cosine-distance}+\alpha], 0\},
\end{aligned}
\end{equation}

Where $N$ is the sum of all possible triplets. 
The collection of all the possible triplets is generated by the output of Cluster-CCA model, it is easy to fulfill the condition defined in Eq.(\ref{triplet}), because the new audio/visual representations have already learned pairwise-based correlation and cluster-based correlation which results in almost pairwise examples of the same class group more closer than the pairwise example from different classes. The triplet loss values of most triplets are zero and these triplets have no contribution to the sum of triplet loss, which lead to the final average of loss values close to zero.  
In particular, when a loss has $||T(i)-S(j)||_{cosine-distance}+\alpha < ||T(i)-S(k)||_{cosine-distance}$, it is equal to zero, the loss has no contribution to optimizing the final loss. Our experiment follows~\cite{hermans2017defense}, a better triplet loss optimization is to ignore all the triplet when its loss is zero, so that the triplet loss can be fast converged and the optimization will be more effective~\cite{schroff2015facenet}. 

It is impossible for us to calculate all the argmin and argmax among all the training dataset. Because in our experiment dataset, we have around 1K examples in MV-10K dataset and more than 2K examples in the VEGAS dataset for each class, which result in a large number of possible triplets. And computation in this way may bring bad generation and over-fitting. 
In this paper, we follow the FaceNet method~\cite{schroff2015facenet} and select triplets to remove all negative/positive samples in a batch when its triplet loss is zero.

\section{Experiments}
\subsection{Dataset and Evaluation Metric}
We evaluate our model on two different video datasets: VEGAS~\cite{zhou2018visual} and MV-10K~\cite{zeng2018audio}. Each video from these datasets contains audio track and visual track, and both data can be represented with high-level features. Our goal is to learn the correlation between this two-view of high-level features. We adopted Mean Average Precision (MAP) and Precision-Recall Curve (PRC) for quantitative performance evaluation.

MV-10K Dataset refers to our previous work~\cite{zeng2018audio}. This dataset is a small subset of large-scale video dataset YouTube-8M\footnote{https://research.google.com/youtube8m/} which contains 10,000 (10k) videos with a "music video" label, and each video ranging from 213 to 219 seconds. Audio and visual track features are extracted in the MV-10K dataset. The YouTube-8M dataset has already released the audio and visual features, respectively extracted by VGGish model trained on Audioset\footnote{https://research.google.com/audioset/}, and pre-trained Inception V3 model trained on ImageNet dataset\footnote{http://www.image-net.org/}.  Based on the frame-level audio features, we applied 10 pre-defined music categories to annotate all videos, we assume these videos have certain knowledge with the music categories and audio-visual pairs shared a single music category.

The VEGAS dataset\cite{zhou2018visual} selected videos from Google Audioset by 10 categories and applied Amazon Mechanical Turk to do data cleaning, the 10 categories are human/animal sounds (chainsaw, helicopter, drum, printers, fireworks, dog, rail transport, baby crying, human snoring, water flowing and rail transport). 
The length of a video ranges from 2 to 10 seconds and the average is 7 seconds, the percentage of all video which ranges from 8 to 10 seconds is above 55\%. In our experiments, we use 28,103 videos to evaluate our architecture.

\textbf{Evaluation Metric}.
In our work, we use MAP and PRC as metrics to leverage our architecture. We focus on category-based cross-modal retrieval, where the system generates a ranked list of documents in one modality by a query in another modality. The samples with the same category as that of the documents are regarded as relevant. Moreover, it takes the location of retrieved document in the rank list into account. The more related documents appear in the top rank list, the higher MAP value it has. 
\subsection{Training Setting}
In our experiments, we set parameters for our deep TNN-C-CCA model as follows.
\begin{enumerate}
\item For deep TNN, there are three branches: anchor branch, positive branch, and negative branch. For each branch, they will go through a full connection. Anchor branch has its own parameters, positive and negative branches share the same parameters.
When taking audio sample as an anchor, the positive and negative are visual samples. We set four hidden layers for each full connection. The number of units per layer is respectively set to 100, 100, 100, 10 for audio branch and 200, 200, 200, 10 for visual branch.
If taking visual as the anchor, the positive and negative samples are from audio samples. We set the number of units per layer for visual branch to 200, 200, 200, 10, and 100, 100, 100, 10 for audio branch. 
\item We set the correlation component for all the following experiments as 10. We set the probability of dropout as 0.2 and use $tanh$ as activation function for each hidden layer and use $sigmoid$ as the activation function in the last layer.
\item We separately divided the training set ranges from 300 to 1,000, and select the best one. The number of training epochs is 20.
\item Our result is the average performance via 5-fold cross-validation. We consider the category balance when we evenly group all the dataset into 5 folds.
\item The Adam optimizer is used for our experiment. The learning rate is set as 0.001.
\end{enumerate}
\subsection{Results on the VEGAS Dataset}
\begin{figure}[t]
\includegraphics[width=6.8cm]{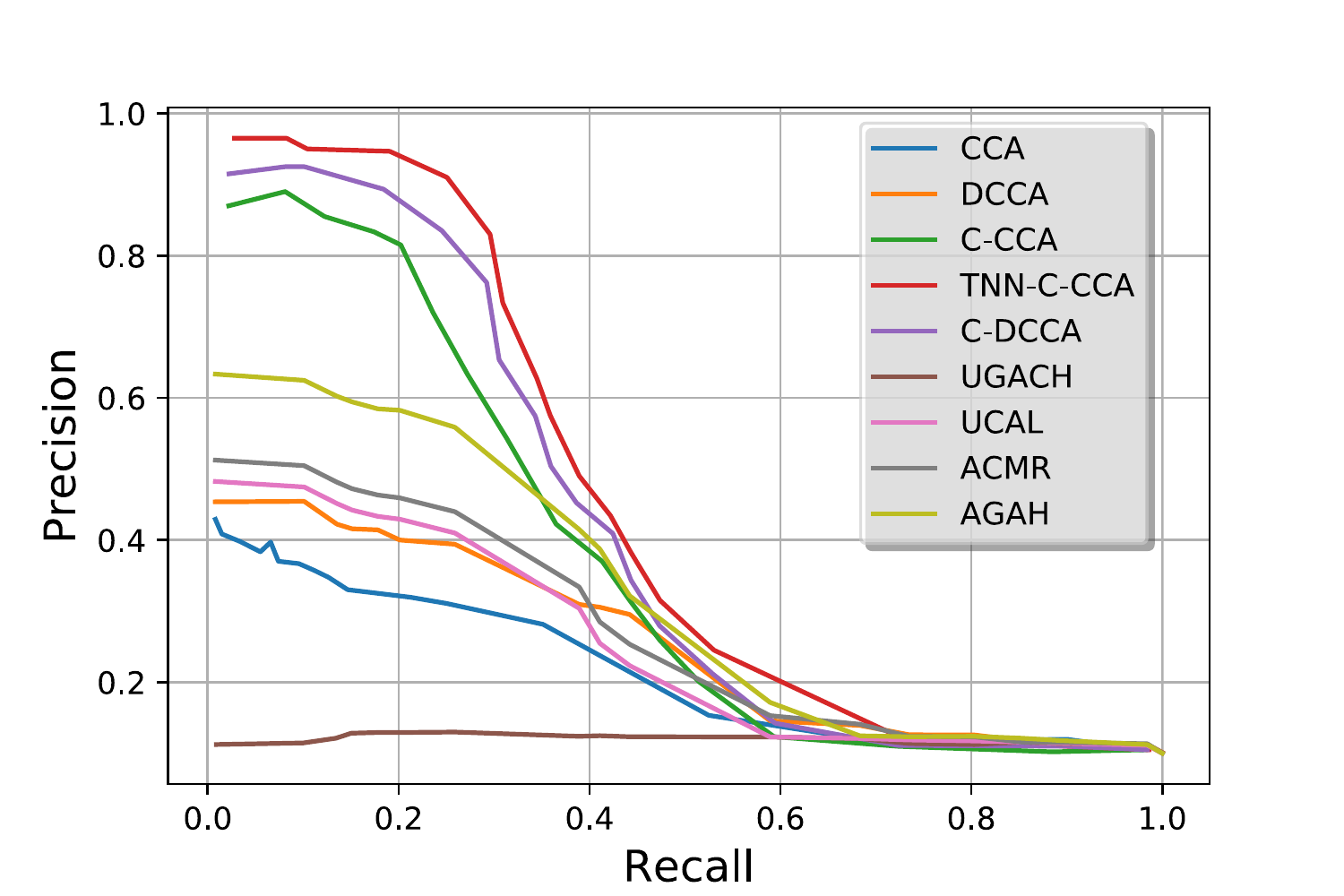}
\includegraphics[width=6.8cm]{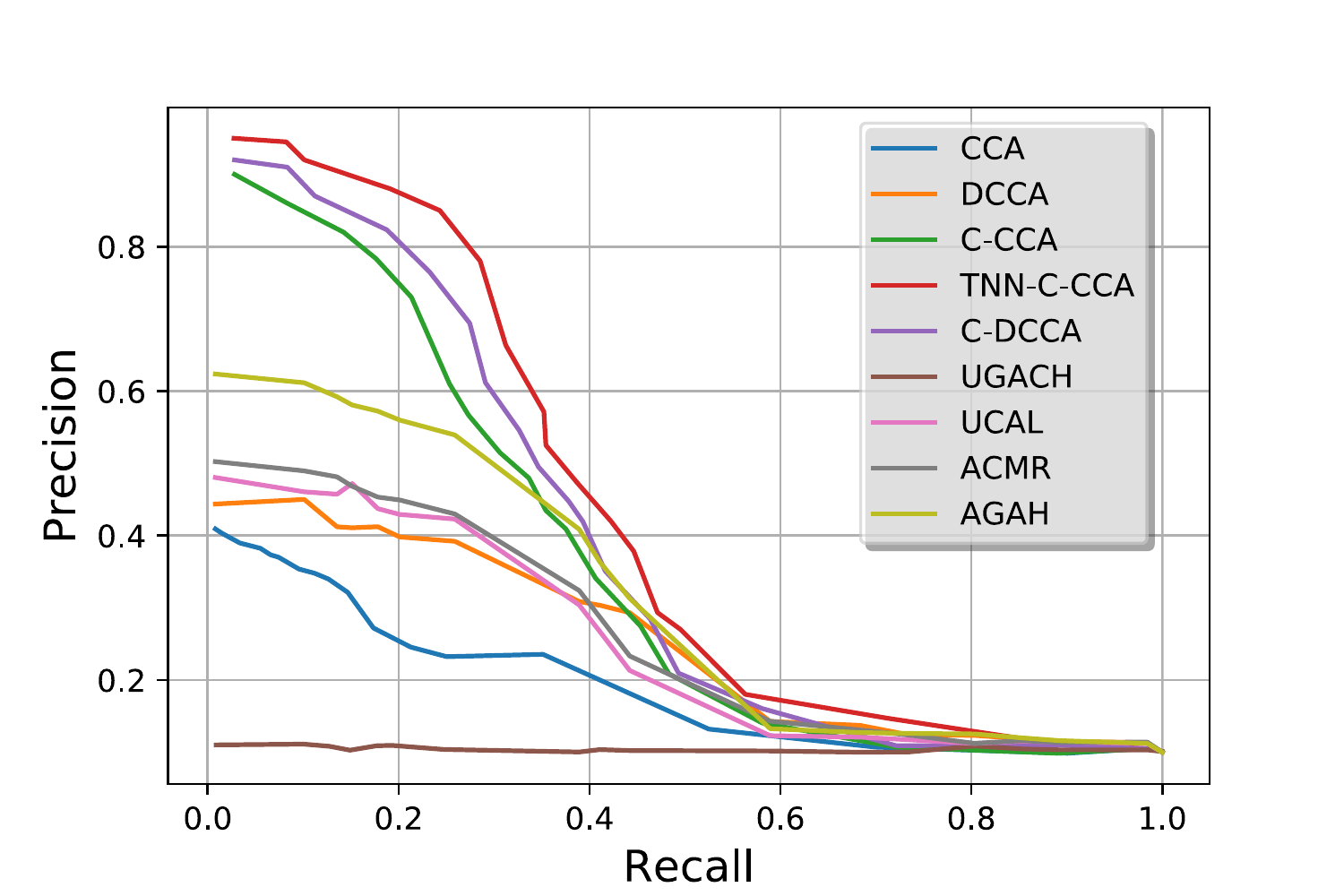}
\centering
\caption{The PRC achieved on the VEGAS dataset with our TNN-C-CCA model and other eight different models. The left figure is for audio-to-visual retrieval, the right figure is for visual-to-audio retrieval.}
\label{fig:vag_avva}
\end{figure}

We report the result of audio-visual cross-modal retrieval task on the VEGAS dataset in the left part of Table~\ref{tab:map_two_datasets} with MAP metric and Fig.~\ref{fig:vag_avva} with PRC. We implement our architecture compared with some existing CCA-variant approaches and non-CCA methods: CCA~\cite{hardoon2004canonical}, DCCA~\cite{andrew2013deep}, KCCA~\cite{akaho2006kernel}, C-CCA~\cite{rasiwasia2014cluster},  C-KCCA~\cite{rasiwasia2014cluster} C-DCCA~\cite{yu2018category}, AGAH~\cite{GuGGLXW19} and etc. as baselines, to show the improvement of our model. For these baselines, we separately implement all of them with the same dimension of outputs and the same parameters. 
\begin{table}[H]
\caption{The MAP scores of cross-modal retrieval between audio and visual contents for our TNN-C-CCA method and some existing state-of-the-art methods on VEGAS dataset and MV-10K dataset.}
\label{tab:map_two_datasets}
\begin{centering}
\begin{tabular}{c|c|c|c|c|c}
\hline
\multirow{2}{*}{\textbf{Models}}
  &\multicolumn{2}{c|}{\textbf{VEGAS Dataset (\%)}}  &\multicolumn{2}{c}{\textbf{MV-10K Dataset (\%)}} \\ 
\cline{2-5}
  & audio2visual
  & visual2audio
  & audio2visual
  & visual2audio
  \tabularnewline
\hline
CCA~\cite{hardoon2004canonical}    &32.43 &32.11 &18.38 &18.17 \\ 
KCCA~\cite{akaho2006kernel}        &28.65 &27.24 &17.81 &17.03 \\
DCCA~\cite{andrew2013deep}         &41.43 &42.15 &18.43 &18.21 \\
C-CCA~\cite{rasiwasia2014cluster}  &65.16 &64.35 &19.71 &19.62 \\
C-KCCA~\cite{rasiwasia2014cluster} &32.41 &32.74 &18.38 &18.11 \\
C-DCCA~\cite{yu2018category}        &70.34 &69.27 &21.79 &20.08 \\
\hline
UGACH~\cite{ZhangPY18}             &17.18 &17.07 &11.11 &11.40 \\
AGAH~\cite{GuGGLXW19}              &57.82 &56.16 &20.74 &20.19 \\
UCAL~\cite{XLYSS17}                &42.68 &41.53 &18.82 &18.47 \\
ACMR~\cite{wang2017adversarial}    &45.46 &43.12 &19.02 &18.63 \\
\hline
LSTM\_C\_CCA                       &66.62 &71.34 &19.11 &18,89 \\
TNN-C-CCA                          &74.66 &73.77 &23.34 &21.32 \\
\hline
\end{tabular}
\end{centering}
\end{table}
According to the experience of our experiments, when the correlation component is set to 10, the CCA-variant approaches can get the best performance\cite{yu2018category, yu2019deep}. Here we use the MAP value as our main performance metric, the MAP of 10 correlation components is much better than the other number of ten multiples correlation components. We set the dimension of outputs of all baselines as 10. The dimensions of the audio feature as inputs are $L*128 (L\in[2, 10])$, the dimensions of visual feature as inputs are $L*1024 (L\in[2, 10])$. For each audio-visual pairwise, $L$ for the audio and the visual are the same. Then feed them into a mean layer to make all the audios and all the visual samples respectively have the same dimensions, to make it possible to calculate the correlation in the shared space with CCA-variant approaches. Especially, the DCCA and the C-DCCA have the same structures of hidden layers. We did all the experiments for each model with 5-fold cross-validation. All models were done by the same structure of folds and the structure established considers balance factor. Each fold contains the same number of samples in each category and 10 categories are kept simultaneously in each fold. 
Table~\ref{tab:map_two_datasets} shows that all CCA variants with category information as training such as C-CCA, C-KCCA, LSTM-C-CCA, and C-DCCA are much better than training without any class as inputs such as CCA, DCCA, and KCCA. The best performance without category information training is DCCA. The MAP of audio-to-visual retrieval is 41.43\% and the MAP of visual-to-audio is 42.15\% over VEGAS dataset, which outperforms the CCA method: the MAP of audio-to-visual retrieval is 32.43\% and the MAP of visual-to-audio retrieval is 32.11\%, and are much better than the KCCA method: the MAP of audio-to-visual retrieval is 28.65\% and the MAP of visual-to-audio is 27.24\%. Compared with the above unsupervised CCA-variant method, the supervised CCA variants can get higher MAP performance. Taking C-CCA as an example, the MAP of audio-to-visual retrieval is 65.16\% which has 23.63\% improvement and the MAP of visual-to-audio retrieval is 64.35\% which has 22.20\% improvement. C-DCCA not only considers the pairwise correlation but also learns the category-based similarity correlation by enlarging the number of pairs if the paired data points have the same category information. In our experiment with this dataset, we establish new possible pairs within the same category for each sample in the train set, then select 50\% pairs for each sample to enlarge the train set. There are three main shortages of C-DCCA method: 1) because it deeply relies on the balance of pairwise correlation and category-based correlation which is adjusted by a hyper-parameter $beta$, it is very hard to set the best $beta$ during the training. 2) when we do model generation for new dataset input, the method can not reduce the noisy pairs which belong to the paired data from other categories closer than the paired data from its category. 3) it is really time-consuming and space-consuming during the training. 

To overcome three shortages, we put forward TNN-C-CCA model with the aim of learning a more reliable correlation in the common space and learning better new joint embeddings for each modality to compute the similarity. Table~\ref{tab:map_tnn_x} shows that our TNN-C-CCA model can get a MAP of 65.62\% for audio-to-visual retrieval and the MAP of 63.30\% for visual-to-audio retrieval by randomly selecting the 150 negative samples for each anchor in the training set. Compared with Cluster-CCA without considering negative information, the Map of visual-to-audio retrieval is improved by around 7\%. However, randomly selecting the negative samples are not statistical reliability, which brings trouble for re-implementing the experiments to get the same result. In theory, we hope to consider all the negative samples, but in fact, for each sample, there almost have 16,800 negative samples and exist $N^{2}$ (N is the size of the training set.) training samples, it is the time- and space- consuming in the case of TNN-C-CCA. In order to balance the time- and space- consuming, and consider the negative samples, according to these works~\cite{schroff2015facenet, hermans2017defense, wu2017sampling}, we build triplets (anchor, positive and negative) inside a batch for training. If the size of the training set is $N$ and the number of the batch is $B$, the batch size is the floor of $N/B$. The samples of all categories balance in each batch. In each batch, there are $\sum_{i=1}^{10} \frac{N_{i}^{2}(N-N_{i})}{B^{3}}$ triplets, and the training set size is $\sum_{i=1}^{10} \frac{N_{i}^{2}(N-N_{i})}{B^{2}}$. Established triplets in a batch, it can save $\sum_{i=1}^{10} \frac{N^{2}}{B^{2}}(N-N_{i})(B^{2}-1)$ training set compared with building all triplets one time, where the $N_{i}$ is the number of pairs with class $i$ in train set. And the performance is much better than that of the C-DCCA and other baselines, the MAP of audio-to-visual retrieval is 74.66\% which has 4.28\% improvement compared to C-DCCA model. The MAP of visual-to-audio retrieval is 73.77\% which has 4.5\% improvement compared to C-DCCA model. In addition, we compare TNN-C-CCA model with four state-of-the-art cross-modal retrieval methods. As shown in Table~\ref{tab:map_two_datasets}, the performance of our TNN-C-CCA model is much better than that of novel adversarial learning methods.
\begin{figure}[t]
\centering
\includegraphics[width=6.9cm]{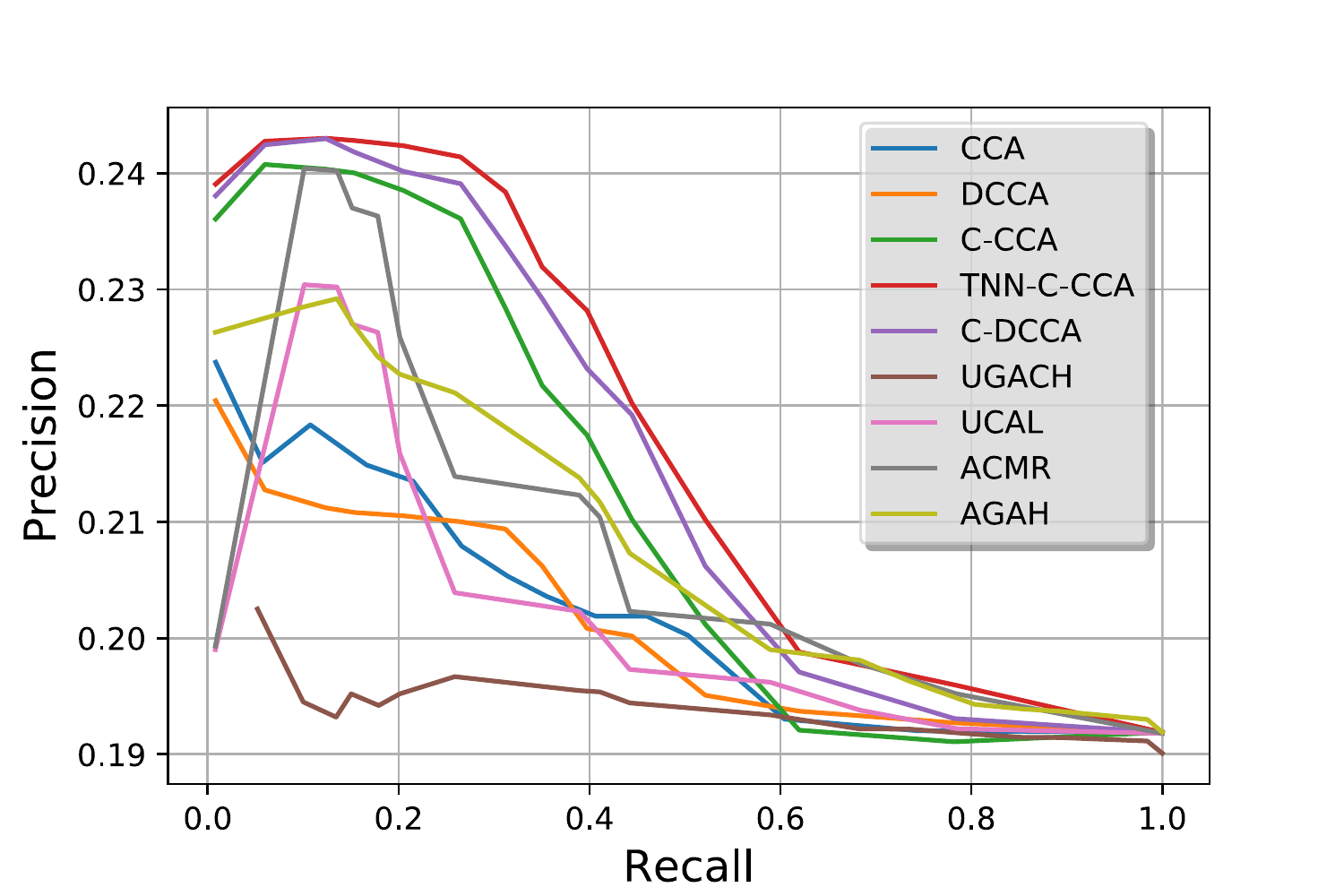}
\includegraphics[width=6.9cm]{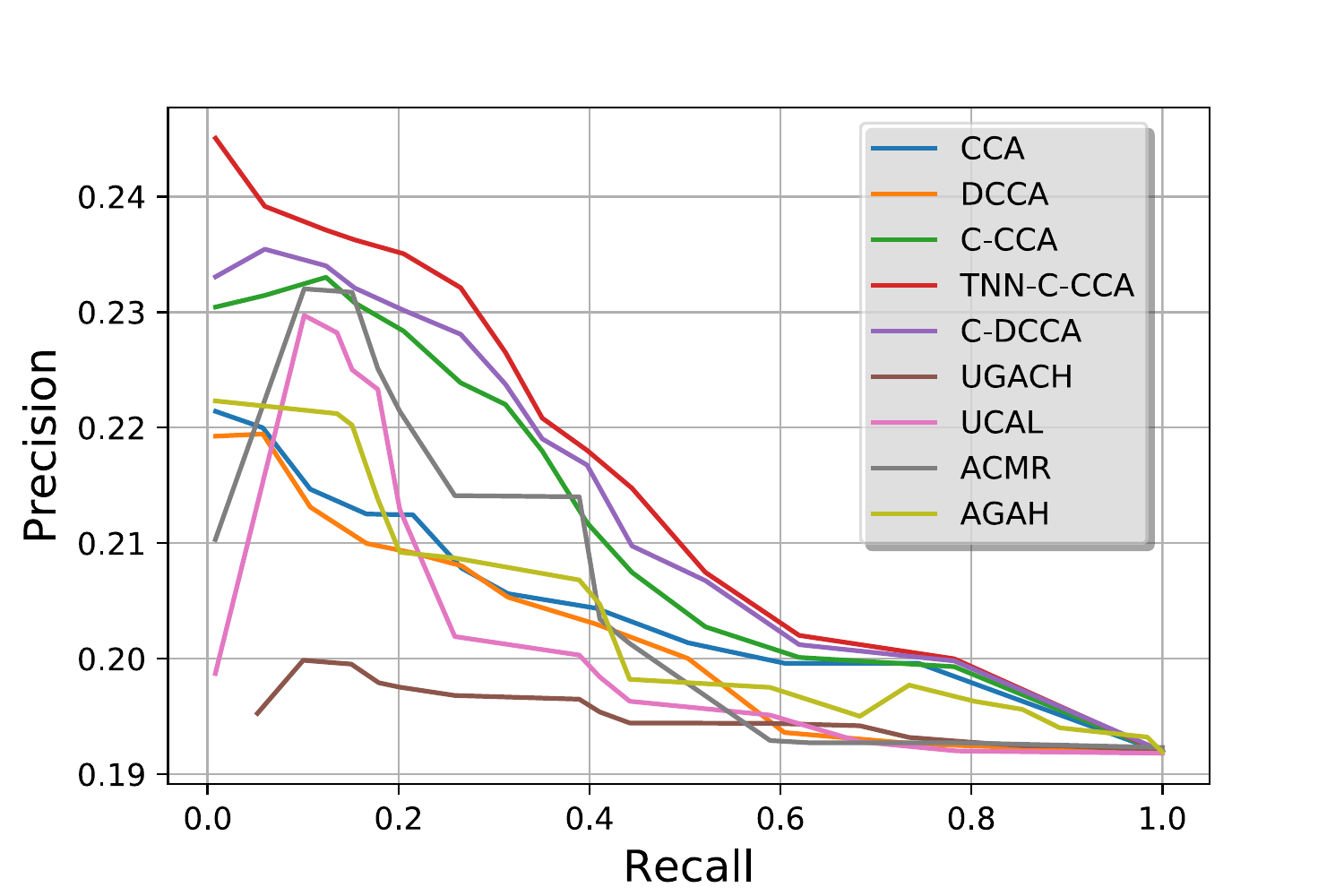}
\caption{The PRC achieved on the MV-10K dataset with nine different models. The left figure is for audio-to-visual retrieval, the right figure is for visual-to-audio retrieval.}
\label{fig:avva}
\end{figure}
\begin{table}[H]
\caption{The MAP scores of audio-visual cross-modal retrieval for our TNN variant methods}
\label{tab:map_tnn_x}
\begin{centering}
\begin{tabular}{c|c|c|c|c|c}
\hline
\multirow{2}{*}{\textbf{Models}}
  &\multicolumn{2}{c|}{\textbf{VEGAS Dataset}}  &\multicolumn{2}{c}{\textbf{MV-10K Dataset}} \\ 
\cline{2-5}
  & Audio2Visual
  & Visual2Audio
  & Audio2Visual
  & Visual2Audio
  \tabularnewline
\hline
C-CCA~\cite{rasiwasia2014cluster}  &65.16 &64.35 &19.71 &19.62 \\
TNN (batch all)             &14.18  &13.44  &13.25  &14.02 \\
TNN (batch semi-hard)       &15.18  &14.22  &14.20  &14.17 \\
TNN (batch hard)            &11.18  &12.20  &12.06  &11.59 \\
TNN-C-CCA (rand)            &65.62  &63.30  &19.23  &18.74 \\
TNN-C-CCA (batch semi-hard) &71.35  &70.23  &20.37  &19.97 \\
TNN-C-CCA (batch hard)      &60.71  &58.39  &19.16  &18.85 \\
TNN-C-CCA (batch all)       &74.66  &73.77  &23.34  &21.32 \\
\hline
\end{tabular}
\end{centering}
\end{table}
\subsection{Results on the MV-10K Dataset}
We report the result of audio-visual cross-modal retrieval on the MV-10K dataset in Table~\ref{tab:map_two_datasets} with MAP metric and Fig.~\ref{fig:avva} with the PRC. We compare our model with some previous models published in~\cite{zeng2018audio}. For those models, where the results of audio-visual retrieval are calculated. Based on the previous works, we use the same input features that are used in all models. In Table~\ref{tab:map_tnn_x}, the TNN-C-CCA (rand) model is achieved by selecting the negative and positive in the training set by random to build the triplet as inputs after obtaining the embedding in the common space with Cluster-CCA method. In the experiment, we randomly select 150 triplets for each sample during the training, as shown in Table~\ref{tab:map_tnn_x}. Because it is very hard to select the triplet for each sample. Since it is time-consuming to use all the possible triplets, we select all the triplets within a batch.  For audio-to-visual retrieval as shown in Table~\ref{tab:map_two_datasets}, our model gets the improvement of 1.55\% for MAP and 1.24\% improved for visual-to-audio retrieval task compared with the state-of-the-art model C-DCCA, and the performance of proposed method is much higher than the state-of-the-art non-CCA models: UGACH, AGAH, UCAL and ACMR model.

In Table~\ref{tab:map_two_datasets}, Fig.~\ref{fig:vag_avva} and Fig.~\ref{fig:avva}, it is easy to notice that the MAP of VEGAS Dataset is much better than that of MV-10K Dataset. Two main reasons are explained as follows.

\begin{enumerate}
\item The supervised cross-modal retrieval deeply depends on the accuracy of the label for the samples. In the MV-10K Dataset, the labels are allocated by the feature similarity. It is hard to guarantee the allocated labels are always correct. There exist many noisy labels in this dataset. However, the VEGAS Dataset is annotated by volunteers and the labels are double-checked. The label can accurately reflect the semantic information in both audio and visual modalities.
\item Moreover, video in the MV-10K Dataset is about 216 seconds while the VEGAS dataset is 10 seconds or less. The input of our model is high-level features, this kind of feature is more effective for the short length of the video in this case. Because high-level semantic features will filter that unimportant information. We use the same dimension to represent those two datasets, in general, which leads to long videos losing more information than short videos.
\end{enumerate}
\subsection{Ablation Study of TNN-C-CCA}

To have a good ablation study, we investigate triplet selection for the inputs of TNN model to see how it influences the performance of TNN-C-CCA architecture. We also study the impact of distance using in triplet loss of TNN-C-CCA. Then, we show the visualization of the learned semantic space and display the visualization of retrieval results according to the given audio query. In addition, we discuss the effect of model parameters.
\begin{figure}[t]
\includegraphics[width=10cm,height=5.5cm]{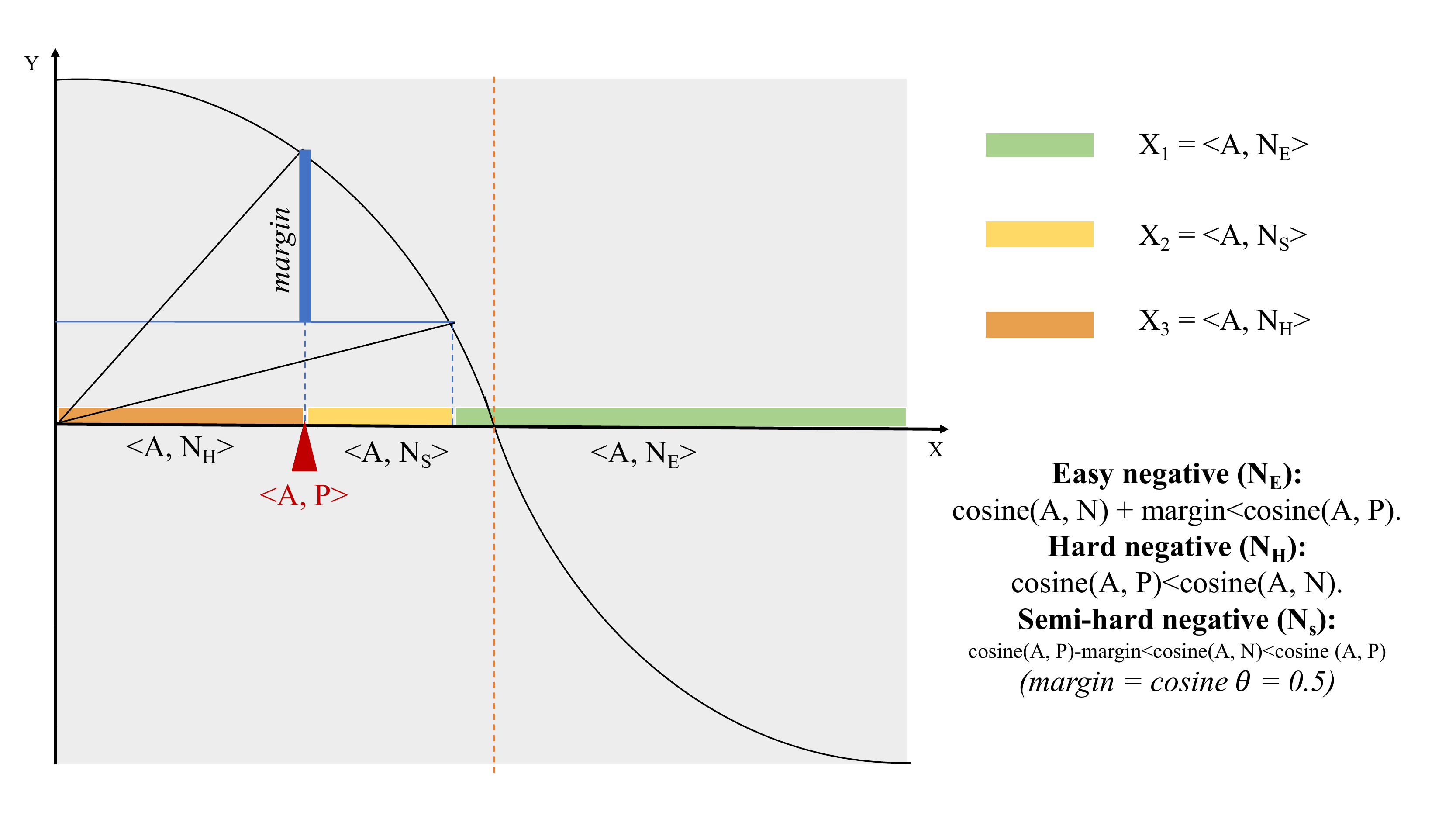}
\centering
\caption{Given an Anchor-Positive pair with its angle <A, P>, those negative samples having the same modality with Anchor as Positive and having different label as Positive, based on the relationship between cosine(A, P) and cosine(A, N), can be classified into three categories: 1) Easy negative, 2) Hard negative and 3) Semi-hard negative.}
\label{triplet_rr}
\end{figure}

\subsubsection{Triplet selection strategies}\mbox{}

According to the relationship between anchor-positive distance and anchor-negative distance, triplets can be divided into three categories. In other words, under the fixed anchor-positive distance, negative samples can be categorized into three classes: easy negative, hard negative and semi negative, as shown in Fig.~\ref{triplet_rr}. During the training, the triplet selection for training TNN-C-CCA model is a very important part. We introduce three triplet selection strategies: batch all when selecting all triplets as training, batch hard when selecting one hard negative-based triplet as training, batch semi-hard when selecting all semi-hard as training.

Table~\ref{tab:map_tnn_x} shows the MAP scores of audio-visual cross-modal retrieval with three triplet selections strategies that are used for training for TNN and TNN-C-CCA.
In TNN model which uses original audio-visual features as input,  batch semi-hard as training can achieve the best performance for audio-visual retrieval. However, in TNN-C-CCA model, batch all can obtain the best performance.

On the other hand, it is obviously that C-CCA with TNN embedding is much better than C-CCA embedding and TNN embedding respectively, the best TNN model (batch semi-hard), which can achieve MAP of 15.18\% for audio-to-visual retrieval and MAP of 14.22\% for visual-to-audio on VEGAS dataset, MAP of 14.02\% for audio-to-visual retrieval and MAP of 14.17\% for visual-to-audio on MV-10K dataset. The TNN-C-CCA (batch hard) can obtain MAP of 60.71\% for audio-to-visual retrieval and MAP of 58.39\% for visual-to-audio on VEGAS dataset, MAP of 19.16\% for audio-to-visual retrieval and MAP of 18.85\% for visual-to-audio on MV-10K dataset. From these results, we can observe that the proposed TNN-C-CCA model gets a significant improvement comparing with C-CCA embedding.

\begin{table}[h!]
  \begin{center}
    \caption{MAP with respect to Euclidean distance and Cosine distance in TNN-C-CCA model}
    \label{tab:distances}
    \begin{tabular}{c|c|c} 
    \hline
      \textbf{Distances} &audio-visual &visual-audio \\
      \hline
      \textbf{Euclidean distance} &0.5300 &0.4206  \\
      \textbf{Cosine distance} &0.7466 & 0.7377  \\
      \hline
    \end{tabular}
  \end{center}
\end{table}

\subsubsection{Distance metrics in triplet loss}\mbox{}

To examine the effectiveness of the distances applied in the triplet loss of TNN-C-CCA model, we briefly introduce the Euclidean distance as follows:
\begin{equation}
\begin{aligned}
||X, Y||_{euclidean-distance} = \sqrt{\sum_{i=1}^{n} (x_{i}-y_{i})^{2}},
\end{aligned}
\end{equation}
where $X=(x_{1}, x_{2}, ..., x_{n})$ and $Y=(y_{1}, y_{2}, ..., y_{n})$ are two points in Euclidean n-space with Cartesian coordinates.

Then, we compared Euclidean distance with Cosine distance in triplet loss of TNN-C-CCA model. Table~\ref{tab:distances} shows the results on VEGAS dataset, which demonstrates Cosine distance is much better than Euclidean distance. In particular, the MAP score is significantly improved. Euclidean distance value is unlimited which may lead to the triplet loss is too large during the training and it is hard to be converged.
\begin{figure}[t]
\includegraphics[width=13.0cm, height=7.5cm]{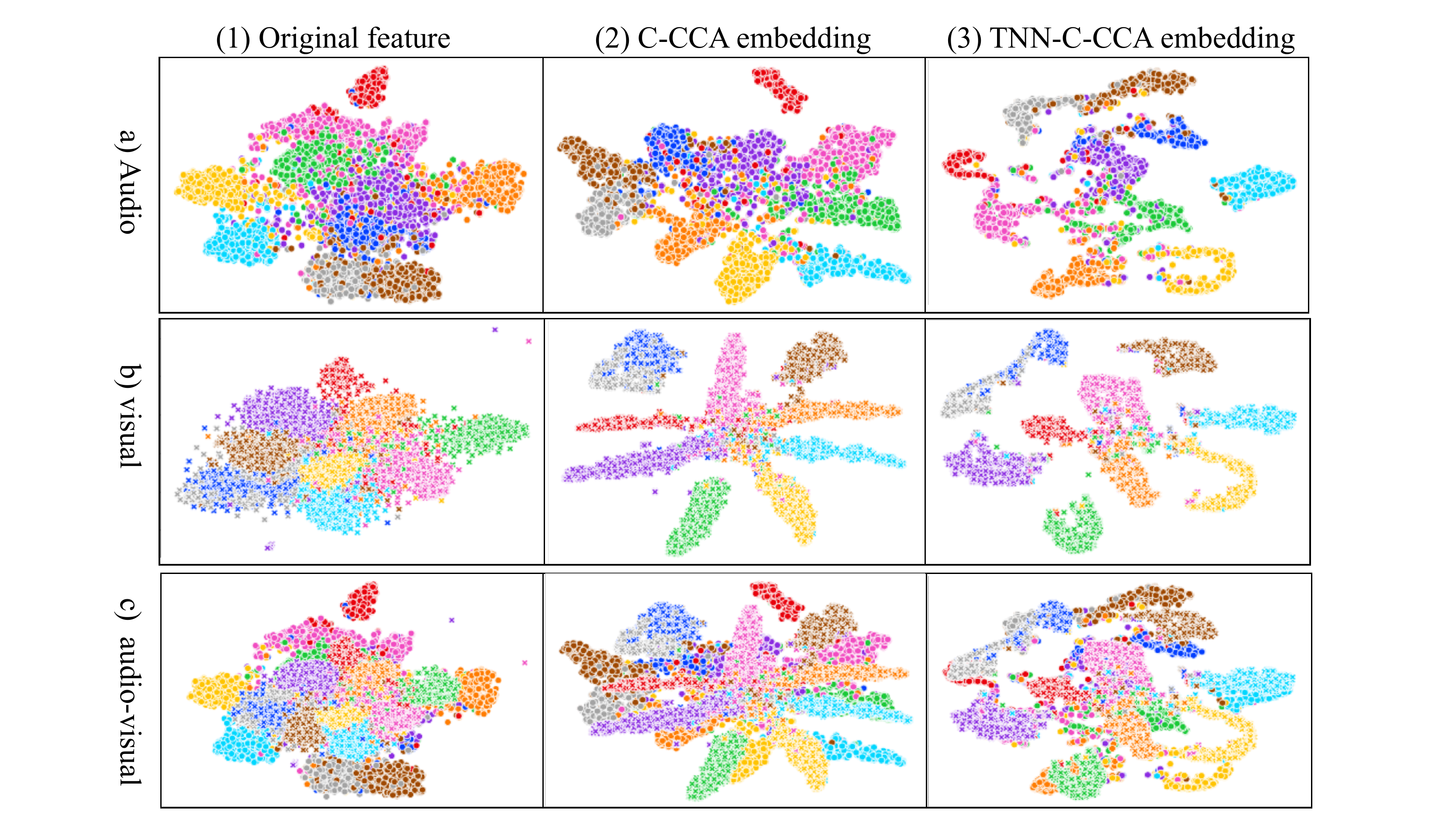}
\centering
\caption{The visualization of the two learned subspace with the t-SNE plot, shows audio, visual and audio-visual in the original feature, C-CCA learning subspace, and TNN-C-CCA learning subspace. The circle sign represents audio, the cross sign represents visual.}
\label{audiovisualspace}
\end{figure}

\subsubsection{Visualization of the learned semantic space}\mbox{}

The goal is to investigate the effectiveness of TNN-C-CCA model combines C-CCA embedding and TNN model on VEGAS dataset. We select one fold as target set with 5,600 samples. The learned common semantic space from C-CCA to generate the semantic features for all samples and then input them into TNN model is to generate more discriminative semantic features by taking negative samples into the training stage. Then, we use t-distributed Stochastic Neighbor Embedding (t-SNE) to implement dimension reduction on the original audio-visual dataset and these features respectively generated from Cluster-CCA and TNN-C-CCA model, where Fig.~\ref{audiovisualspace} shows audio, visual and audio-visual of their raw features, C-CCA features and TNN-C-CCA features. We can see that in Fig. ~\ref{audiovisualspace}, many samples in each category of two modalities scatter and hardly separated, while C-CCA embedding groups into clusters and each cluster represents one category, however, the clusters are not completely discriminative. In the center of space, some samples from different clusters are intersection and hard to be segregated. TNN-C-CCA embedding is much better than C-CCA embedding, those new clusters are more discriminative and samples belonging to the same category are in the same cluster. It indicates that TNN-C-CCA embedding learning effectively improves the performance compared with C-CCA embedding learning.

\begin{figure}[t]
\includegraphics[width=14cm, height=7.8cm]{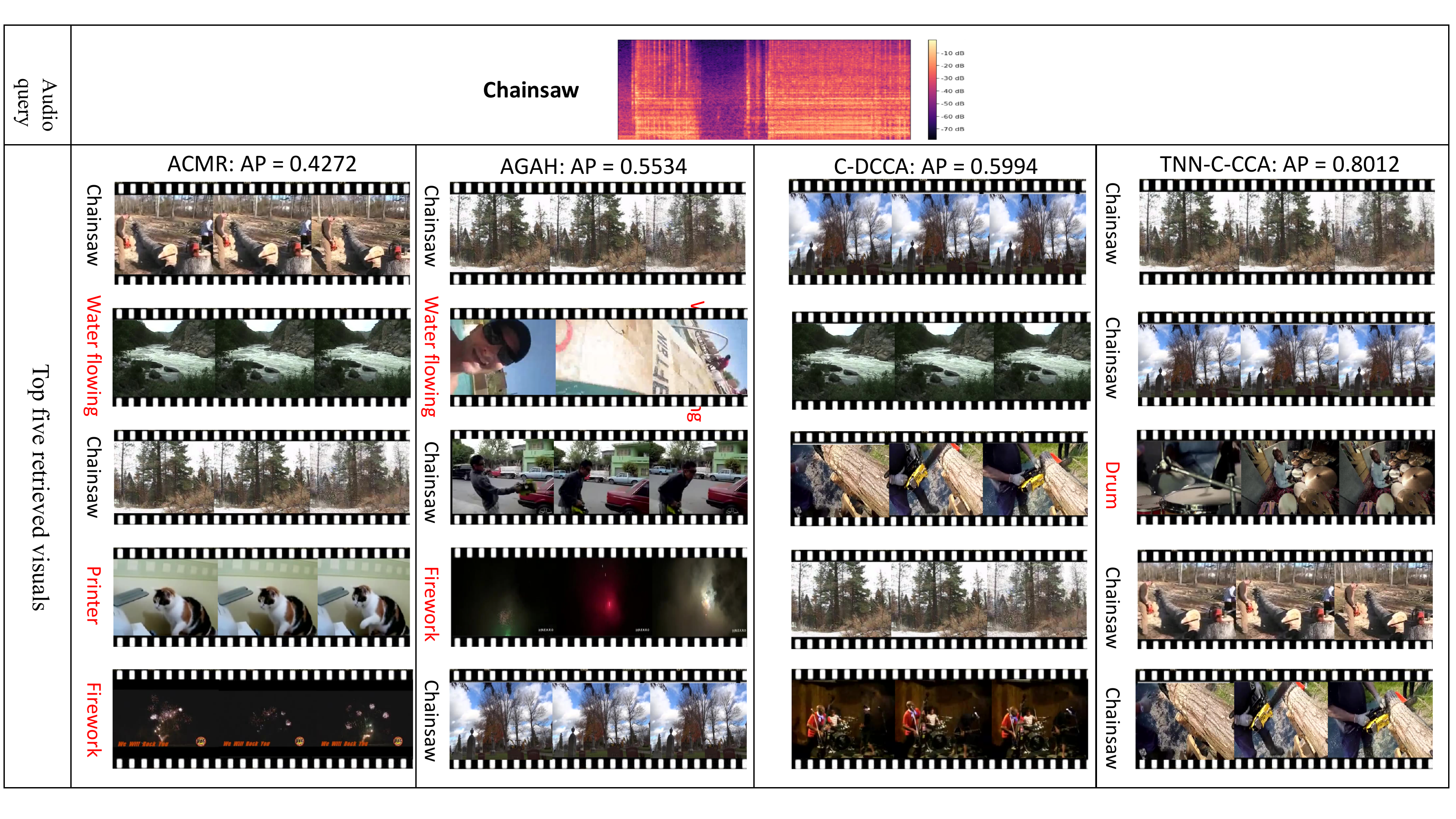}
\centering
\caption{The visualized audio-visual retrieval results of our proposed method and other three best methods, the Cluster-CCA, the AGAH, and the ACMR model. Given an audio as query, the figure shows the top five retrieved visuals.}
\label{resultvisuals}
\end{figure}

Furthermore, we investigate the effectiveness of learned semantic space by the audio-visual retrieval task. We try to compare the retrieval results of our model with the other three best models. Fig.~\ref{resultvisuals} provides audio-to-visual retrieval examples generated respectively by ACMR, AGAH, C-DCCA, and our TNN-C-CCA model on VEGAS dataset for given audio with the "Chainsaw" label as the query. we can observe that the matched top 5 visuals by our TNN-C-CCA is 80\% related to the label "Chainsaw" and average precision (AP) is 80.12\% in all rank lists. For other models, ACMR model is 40\% related to the query label and AP is 42.72\% in all rank list; AGAH model is 60\% related to the query label and AP is 55.34\% in all rank list; C-DCCA model is 60\% related to the query label and AP is 59.94\% in all rank list.

\subsubsection{Effect of model parameters}\mbox{}

In the deep TNN part, batch size and margin play a leading role in the impact of the performance and time-consuming of the system. In this work, we respectively do some experiments on VEGAS dataset to evaluate the impact of batch size and margin.
\begin{table}[h!]
  \begin{center}
    \caption{MAP with respect to different margins with TNN-C-CCA model when batch\_num is 500}
    \label{tab:margin}
    \begin{tabular}{c|c|c|c|c|c|c|c|c|c|c|c} 
    \hline
      \textbf{Margin} &0.1 &0.2 &0.3 &0.4 &0.5 &0.6 &0.7 &0.8 &0.9 &1.0 &1.1 \\
      \hline
      \textbf{audio-visual} &64.73 &68.82 &74.30 &74.59 &75.31 &74.17 &74.15 &73.80 &74.68 &65.30 & 61.28 
      \\
      \textbf{visual-audio} &64.36 & 67.29 &72.45 &73.20 &73.26 &72.42 &72.36 &72.12  &73.04 &62.96 & 58.47 
      \\
      \hline
    \end{tabular}
  \end{center}
\end{table}
\paragraph{\textbf{Margin}}~\cite{cortes1995support} is a region which is bounded by two hyper-planes in the support-vector machines (SVM), when selecting two hyper-planes to split two categories of data. The goal of SVM optimal is to maximize the margin between the vectors of the two categories. The margin of deep TNN is quite similar to the margin in SVM.

\begin{figure}[h]
\centering
\includegraphics[width=6.9cm]{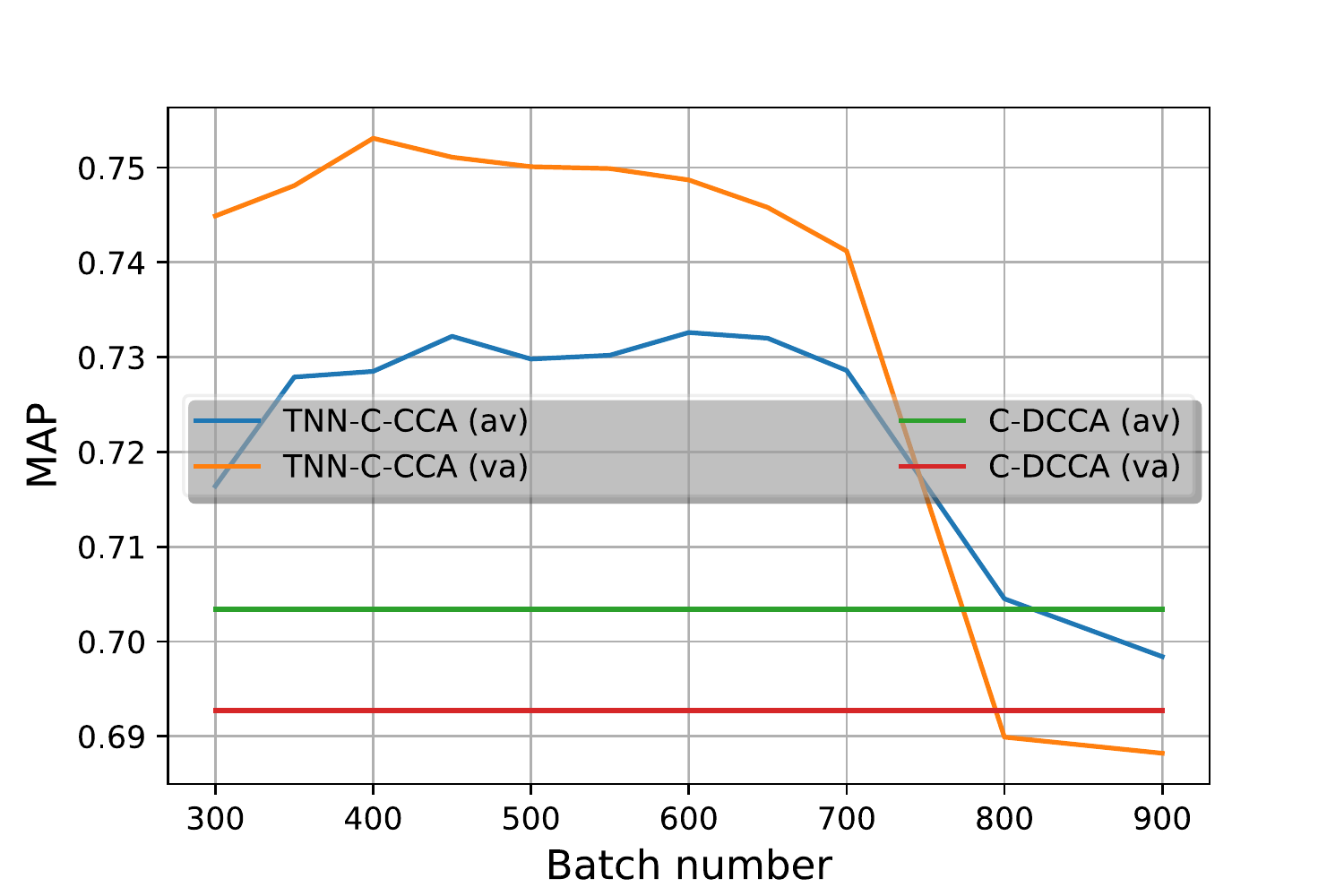}
\includegraphics[width=6.9cm]{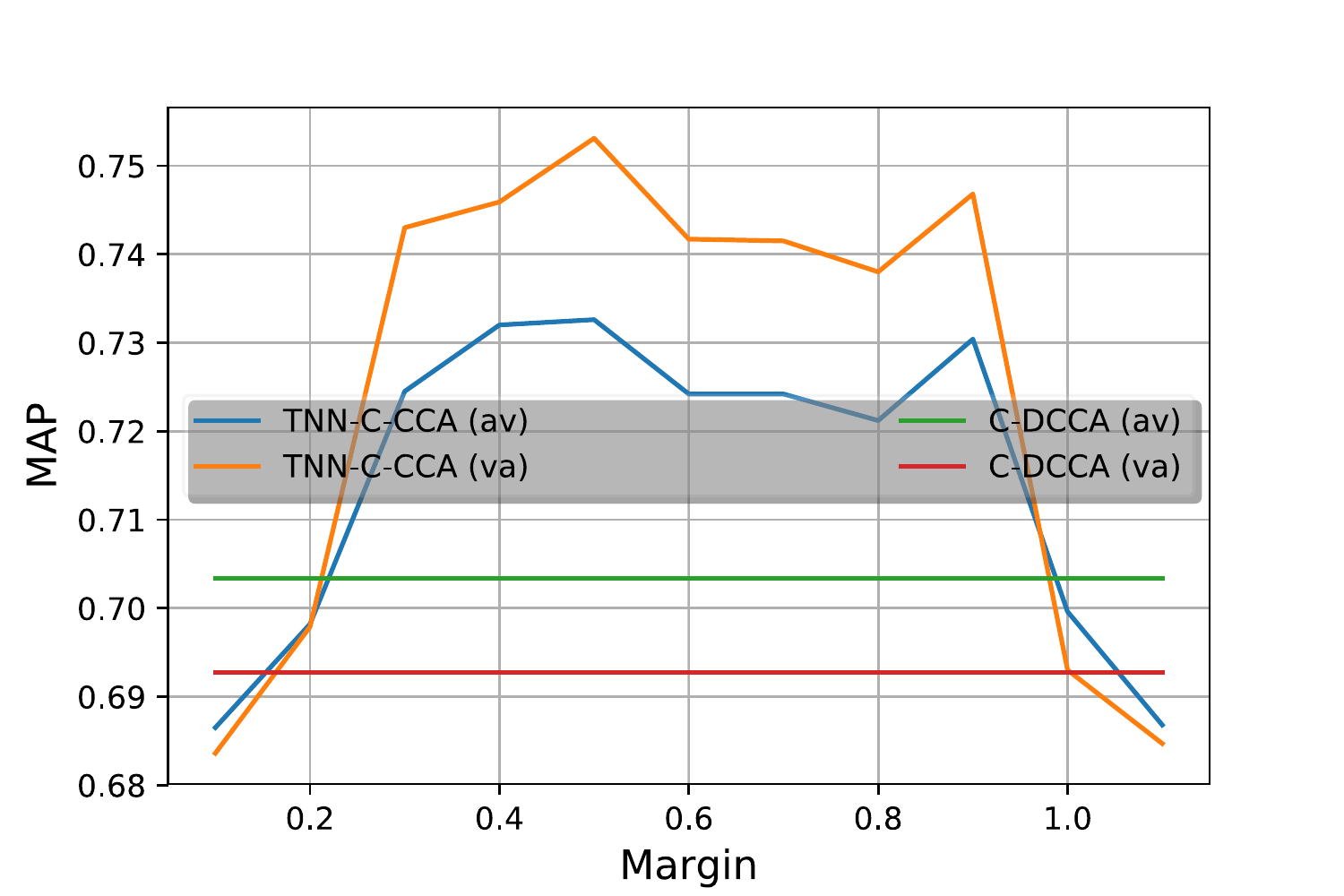}
\caption{The left figure is the MAP curve of TNN-C-CCA and C-DCCA on batch number range from 300 to 700 and the margin are 0.5. The right figure is the MAP curve of TNN-C-CCA and C-DCCA on margin range from 0.3 to 1.0 and the batch number is 500.}
\label{fig:map_batch_margin}
\end{figure}

In our work, we use Cosine distance to calculate the difference among anchor, positive and negative samples, according to our loss function of deep TNN, the effective margin ranges from 0.0 to 2.0. In our experiments, we show the MAP of audio-to-visual retrieval and visual-to-audio retrieval based on the margin ranges from 0.1 to 1.1 by a step as 0.1 and set the number of batches to 500. All the results are listed in Table~\ref{tab:margin}. In order to show the change of MAP values more obviously, we draw the MAP curve based on changing the margin. The right of Fig.~\ref{fig:map_batch_margin} presents when the margin range from 0.3 to 0.9 by step as 0.1, the MAP value has no big change. When the margin is 0.5 the MAP can get the best performance. As margin increases from 0.1 to 0.5, the MAP increases from 64.73\% to 75.31\% for audio-to-visual retrieval and from 64.36\% to 73.26\% for visual-to-audio retrieval. While the margin ranges from 0.5 to 1.1, the MAP decreases from 75.31\% to 61.28\% for audio-to-visual retrieval and from 73.26\% to 58.47\% for visual-to-audio retrieval. 

\paragraph{\textbf{Batch size}} is a hyper-parameter in machine learning, which defines the numbers of samples to update the model weights in one iteration. The number of batches is the number of iterations used in the experiment. Generally, the training dataset can be divided into one or more batches. In our experiments, we defined different batch sizes by changing the number of batches. We divided our training set into different batches ranging from 300 to 900 by a step as 50.
\begin{table}[h!]
  \begin{center}
    \caption{MAP in respect to different batch sizes with TNN-C-CCA model when margin is 0.5}
    \label{tab:batch_time}
    \begin{tabular}{c|c|c|c|c|c|c|c|c|c|c|c} 
    \hline
      \textbf{Batches} &300 &350 &400 &450 &500 &550 &600 &650 &700 &800 &900 \\
      \hline
      \textbf{a-v} &74.49 &73.63 &75.31 &74.50 &74.51 &74.99 &74.87 &74.58 &74.12 &62.96 &61.28
      \\
      \textbf{v-a}  &73.16 &71.47 &73.26 &72.55 &72.98 &73.22 &72.85 &72.79 &71.64 &65.30 &58.47\\
      \hline
      \textbf{Time(h)} &32 &27 &21 &16 &12 &9 &6 &4 &3 &2 &2\\ 
      \hline
    \end{tabular}
  \end{center}
\end{table}
Table~\ref{tab:batch_time} shows the MAP and time-consuming (hour) of audio-to-visual retrieval and visual-to-audio retrieval. CCA, KCCA, C-CCA, DCCA, and C-DCCA will take about 2, 3, 3, 4 and 7 hours respectively. In general, time-consuming take more time, the performance will be better. When the number of a batch is 400, the batch size is about 55 (batch size=training set/batch number), which can get the best MAP value of 75.31\% for audio-to-visual retrieval and 73.26\% for visual-to-audio retrieval compared with other number of a batch. Overall, the MAP value has no big difference when the number of batch ranges from 300 to 700. The big difference of running time in audio-visual cross-modal retrieval is when the number of a batch is 300 and the samples in the batch are balanced, it needs almost 32 hours to finish the experiment. There are around 70 samples in the batch, including 63 negative samples and 6 positive samples combination, totally in the batch there are 6*63*70=264640 triplets. 
When the training set is divided into 700 batches, the batch size is about 30. In the same situation, in the batch, there are 2*27*30=1620 triplets, it saves more time compared with 300 batches, only taking 3 hours. When the number of batches is set to 800, the MAP will decrease a lot and the performance is close to that of the C-DCCA model. When the batch number is 900, the MAP will be lower than that of the C-DCCA model. In the left of Fig.~\ref{fig:map_batch_margin}, the top MAP is 400 batches. In the left part of the curve, as the batches increase from 300 to 400, the MAP will get a bit larger. In the left part of the curve, the number of a batch from 500 to 900, the MAP is degraded. When the number of batches reaches 800, our model gets the same performance as C-DCCA. When the number of a batch is smaller than 800, it will get lower than that of C-DCCA. 

The above experiment results show that our model can outperform other methods when we set effective parameters (margin and batch size).
We respectively do the experiments based on one of them as the main variable. There are a lot of combinations between batch size and margin. In our experiments, we fixed the margin as 0.5 and make the batch size as a variable. Better batch size is obtained based on better MAP. Secondly, when batch size is fixed and the margin is made as a variable, we can get a better margin.

\begin{figure}[t]
\includegraphics[width=6.8cm,height=5cm]{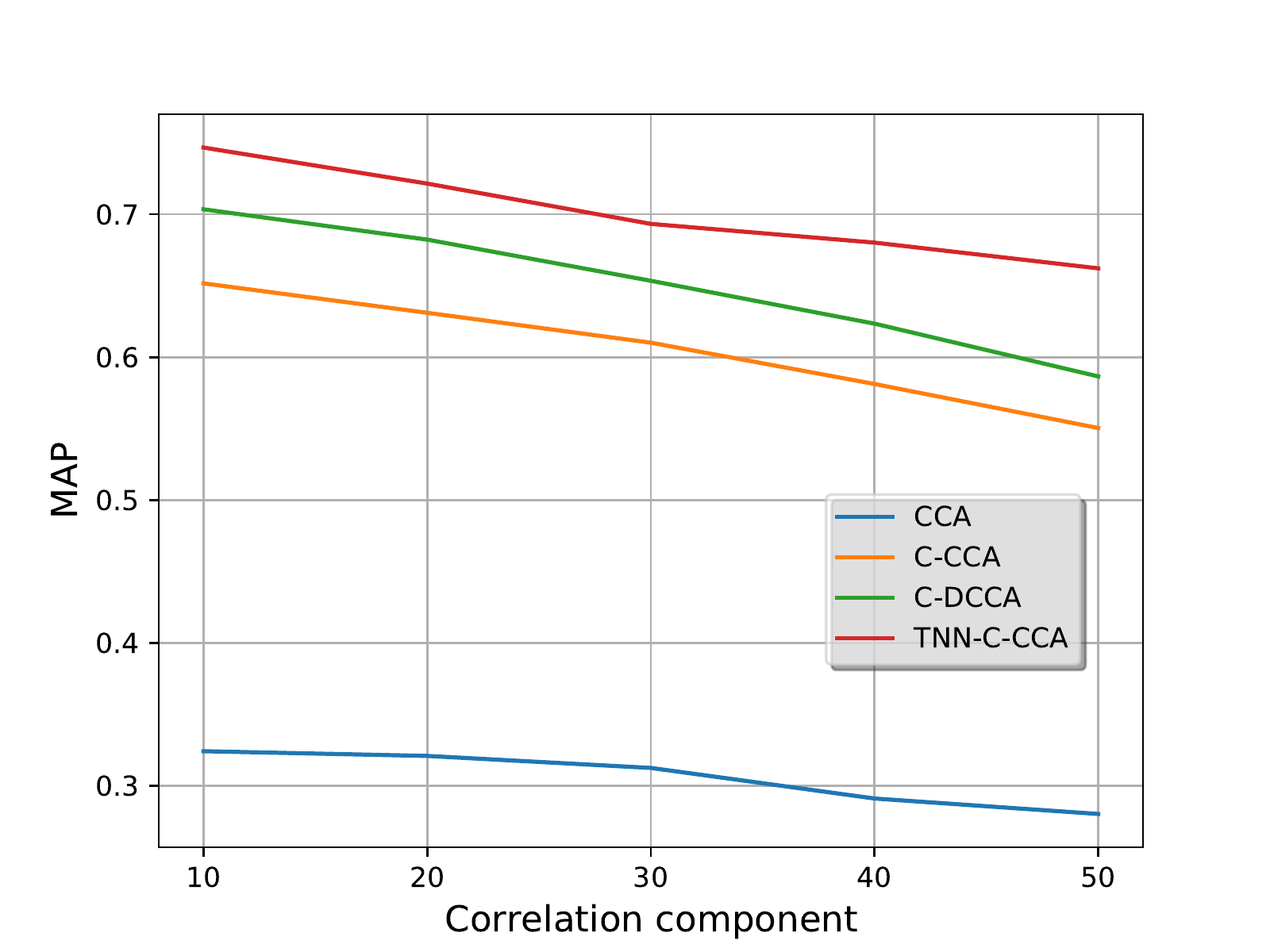}
\includegraphics[width=6.8cm,height=5cm]{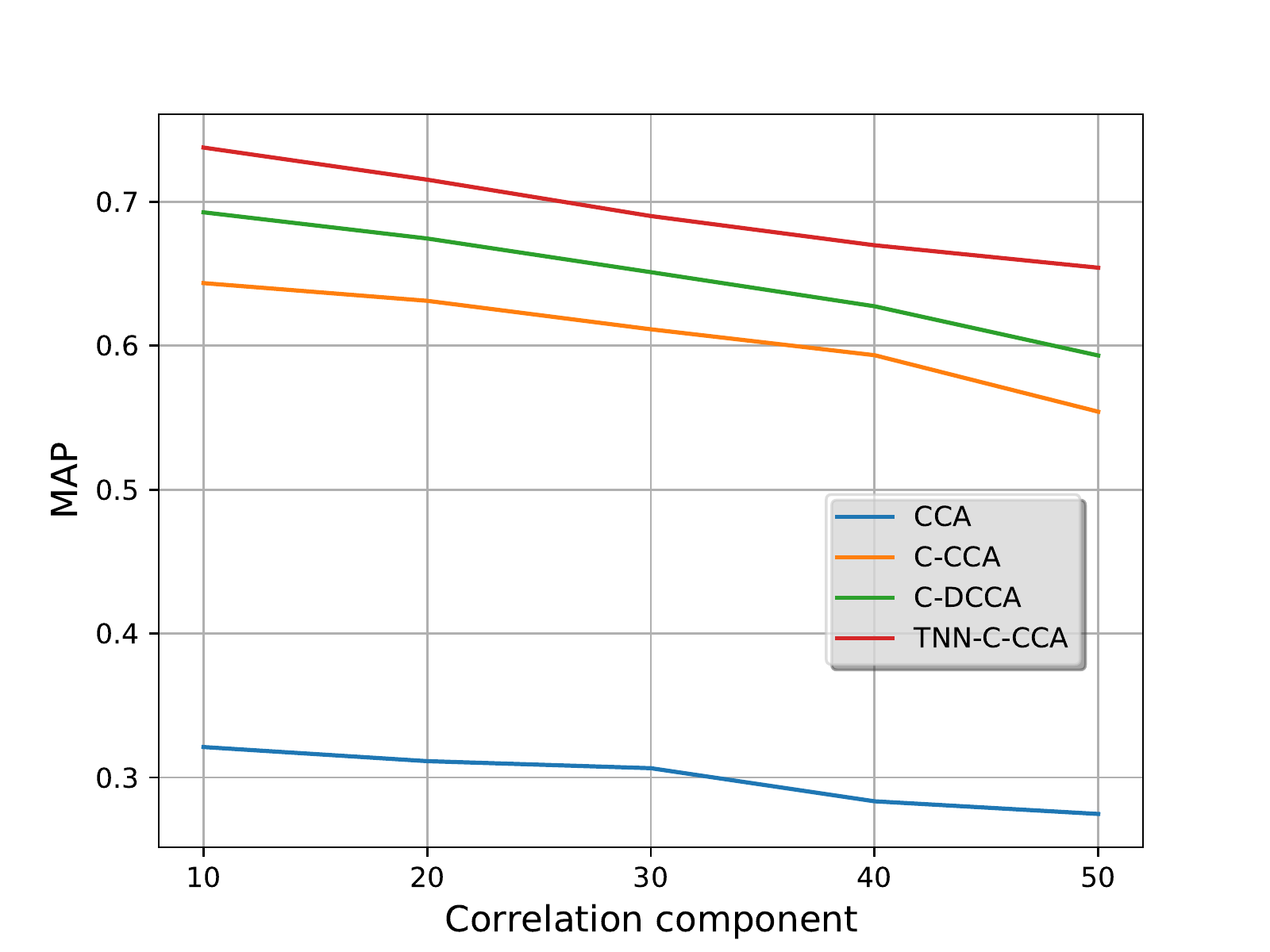}
\centering
\caption{The MAP curve of the correlation component changes from 10 to 50, the corner point in the curve represents the correlation component of X-axis and MAP of Y-axis, which use line to connect two adjacent points.The left part is audio-to-visual retrieval and the right part is visual-to-audio retrieval.}
\label{map_component}
\end{figure}
\paragraph{\textbf{Correlation components}} In addition, the number of correlation components in the CCA-variant method are very important, in order to investigate the correlation structure of learned representation among the four approaches.  Fig.\ref{map_component} shows the MAP curve based on the change of the number of components for all the four models. In our experiments, as for our architecture TNN-C-CCA, the dimension of Cluster-CCA and the dimension of output in deep TNN are the same. It is very clear that the number of correlation components is set to 10 which can achieve the best MAP 74.66\% for audio-to-visual retrieval and 73.77\% for visual-to-audio. As the component decreases, the performance will go down. Especially, it is not a big change in the CCA paradigm at 10, 20, 30, but with the decrease at 40 and 50.

\label{sec:Conclusion}
\section{Conclusion}
In this work, we propose a new deep architecture that consists of Cluster-CCA and deep TNN model. Our architecture can get both benefits of the Cluster-CCA and deep TNN such that completely consider the suitable location of each data point in the shared subspace based on the pairwise correlation and semantic label allocation. The deep TNN model is a supplement of Cluster-CCA model by learning the similarity distance between all pairs within the same class and compares the similarity distance with all possible pairs cross different views. This can help to learn more discriminative embedding space between audio and visual. We applied two different audio-visual datasets to evaluate the performance of our architecture with the PRC and MAP metrics. Audio and visual features are respectively represented by the advanced pre-trained deep CNN based feature extractors for both datasets. The result of the experiments proved that our model can outperform other state-of-the-art cross-modal retrieval models. In order to further investigate the capability of cross-modal embedding learning, we design more extensive experiments for ablation studies where triplet selection strategies, distance metrics, visualization of learned semantic space, and effect of model parameters are investigated.

In the future, we would like to extend our model to support retrieval across other different multi-modalities, such as image-text, audio-text, and video-text cross-modal retrieval. We would like to explore generative adversarial networks (GAN) methods to improve our architecture, and we attempt to extend our current framework to achieve unsupervised cross-modal retrieval to solve the problem of weekly annotated labels like our MV-10K dataset.

\section*{Acknowledgements}
This work was supported by JSPS Grant-in-Aid for Scientific Research (C) under Grant No. 19K11987.
\bibliography{sample-bibliography.bib}
\bibliographystyle{ACM-Reference-Format}

\end{document}